\pdfoutput=1
\documentclass[foundations,article,submit,pdftex,moreauthors]{mdpi}

\firstpage{1}
\pubvolume{1}
\issuenum{1}
\articlenumber{0}
\pubyear{2026}
\copyrightyear{2026}
\datereceived{ }
\daterevised{ }
\dateaccepted{ }
\datepublished{ }

\usepackage{mathtools}
\let\linenumbers\relax

\Title{Cosmic Heterogeneity Shapes the Evolution of Dark Energy and Time}

\Author{Jacky David Yang $^{1,2,}$* and Hans-Otto Carmesin $^{2,3,4}$}

\AuthorNames{Jacky David Yang and Hans-Otto Carmesin}

\address{%
$^{1}$ \quad Altes Gymnasium Bremen, Kleine Helle 7/8, 28195 Bremen, Germany\\
$^{2}$ \quad University of Bremen, Otto-Hahn-Allee NW\,1, 28359 Bremen, Germany; carmesi1@uni-bremen.de (H.-O.C.)\\
$^{3}$ \quad Studienseminar Stade, Bahnhofstr.~5, 21682 Stade, Germany\\
$^{4}$ \quad Gymnasium Athenaeum and Observatory Stade, Harsefelder Str.~40, 21680 Stade, Germany}

\corres{Correspondence: jackydavid.yang@gmail.com}

\abstract{Early- and late-Universe measurements of the Hubble constant disagree, a puzzle known as the Hubble tension. We derive cosmological parameters from first principles, using only the dynamics of cosmic volume and the possibility that matter can be unevenly distributed. Within this framework, three of the six parameters of the standard model are obtained from the underlying dynamics rather than fitted to observations: volume formation determines the dark-energy density parameter to be two thirds, spatial curvature vanishes, and the matter density follows from the remaining cosmic energy budget. The Hubble constant remains an observational input because it represents a calendar date rather than an intrinsic property of the Universe, while the two remaining parameters are likewise taken from observation. Once cosmic unevenness is included as a dynamical component, both the expansion rate and the dark-energy density parameter become redshift dependent. The latter increases by eight percent toward the present, from 0.679 in the homogeneous early Universe to 0.735 today, consistent with independent probes spanning several orders of magnitude in redshift, including the microwave background, supernovae, baryon acoustic oscillations, X-ray clusters and gravitational waves. This same mechanism yields the time-evolving dark energy favored by recent combined analyses without introducing a new free parameter. Integrating the resulting expansion history gives a cosmic age of 13.521~Gyr, shorter than the homogeneous value by 0.306~Gyr. This difference is a global time dilation: dark energy and time evolve together, shaped by the unevenness of the cosmos.}


\begin{document}

\section{The Hubble Tension and the Dynamics of the Universe}\label{sec:intro}

The Hubble tension, the discrepancy between early- and late-Universe determinations of the Hubble constant $H_0$, has become one of the central problems of contemporary cosmology. Its implications reach beyond a single parameter: if $H_0$ is not strictly constant, the Hubble time $t_H = H_0^{-1}$, which underlies the inferred age and temporal evolution of the Universe, must also be reconsidered. Because $H_0$ is moreover tied to the expansion dynamics, and hence to the dark-energy density parameter $\Omega_{\Lambda}$, any reassessment of $H_0$ entails a parallel re-evaluation of dark energy.

The present work re-examines cosmic time and dark energy jointly, based on the volume dynamics (VD) of space \cite{Carmesin2021a,Carmesin2026}. A central feature of the VD is that it does not require the real Universe to be exactly homogeneous: the homogeneity assumed in the standard $\Lambda$CDM model does not hold on all scales \cite{Zwicky1933,Peebles1973}, so that quantities inferred under that assumption, including the commonly quoted cosmic age of $\approx 13.8$\,Gyr, are idealized. The dynamical influence of such inhomogeneity on cosmological averages has been studied extensively within general relativity: spatial averaging of inhomogeneous dust cosmologies yields kinematical backreaction terms \cite{Buchert2000,BuchertRasanen2012}, although the size of the effect for standard perturbations has been debated \cite{GreenWald2014}, and local density fluctuations alone appear insufficient to resolve the Hubble tension \cite{Kenworthy2019}. Conceptually closest to the present work, the timescape scenario attributes differences between inferred expansion rates and cosmic ages to position-dependent clock rates in an inhomogeneous Universe \cite{Wiltshire2007}. The volume dynamics differs from these approaches in one decisive respect: because the fundamental volume portions are quantized, the volume formation driven by heterogeneity does not average to zero; the vanishing mean field $\langle \vec{G}^{*} \rangle$ is accompanied by a non-zero mean square $\langle (\vec{G}^{*})^{2} \rangle$, so that a first-order effect remains where perturbative averaging yields only a small one (see the derivation below and Appendix~\ref{app:derivations}). We show that, once heterogeneity is included as a dynamical component, both the dark-energy density parameter and the progression of cosmic time acquire a redshift dependence.

\section{The Homogeneous \texorpdfstring{$\Lambda$}{Lambda}CDM Model}\label{sec:homogeneous}

Only those elements of the homogeneous $\Lambda$CDM model that are needed below are recalled here; the complete model, the linear-growth treatment and the detailed calculation summarized in this section are given in Appendix~\ref{app:lcdm}. In $\Lambda$CDM, dark energy is modeled as a cosmological constant $\Lambda$ with a constant density $\rho_{\Lambda} = \Lambda c^{2}/(8\pi G)$, entering the first Friedmann--Lemaître equation \cite{Hobson2006}
\begin{linenomath}
\begin{equation}\label{eq:friedmann}
\left(\frac{\dot{a}}{a}\right)^{2} = \frac{8\pi G}{3}\,\rho - \frac{kc^{2}}{a^{2}} + \frac{\Lambda c^{2}}{3}.
\end{equation}
\end{linenomath}
For a spatially flat universe ($k = 0$) with the Planck parameters \cite{Planck2020}, table 2, column 7 (averaging values observed via CMB, gravitational lensing and baryon acoustic oscillations), this yields the familiar homogeneous present-day age $t_{0,\mathrm{hom}} = 13.775$\,Gyr. Because the real Universe is heterogeneous, its effective expansion rate differs from the homogeneous case. A physically consistent treatment, in which $H_0(z)$ is integrated over the expansion history, is developed below.

\section{First-Principles Derivation of \texorpdfstring{$\Omega_{\Lambda,\mathrm{het}}(z)$}{Omega\_Lambda,het(z)}}\label{sec:derivation}

Rather than postulating a redshift dependence of dark energy, we derive it from the volume dynamics (VD) of space, together with the redshift-dependent Hubble constant $H_{0,\mathrm{het}}(z)$ on which it rests. The resulting time evolution of the dark-energy density parameter $\Omega_{\Lambda,\mathrm{het}}(z)$ and of the global (equivalently universal) time difference $\Delta t_{U}(z)$ is founded both on observation and on theory. The full chain of reasoning, leading from a small set of very reliable observations to the functions $H_0(z)$, $\Omega_{\Lambda,\mathrm{het}}(z)$ and $\Delta t_{U}(z)$, is summarized in Figure~\ref{fig:map} (Appendix~\ref{app:map}); the step-by-step derivation is given in Appendix~\ref{app:derivations}. It proceeds from the relativistic energy in the local adequate coordinate system \cite{Carmesin2026,PDG2024} and from the stretching of space near a mass \cite{Schwarzschild1916} to the indivisible, massless volume portions \cite{Carmesin2026}, whose dynamics implies the Schrödinger equation \cite{Carmesin2026,Schrodinger1926} and the remaining quantum postulates \cite{Higgs1964,Hilbert1928,Kumar2018}, as well as gravity and, with the equivalence principle \cite{Carmesin2026,Galilei1638,Einstein1911}, the curvature of spacetime.

Space is a stochastic average of indivisible fundamental volume portions $\delta V$, whose relative additional volume $\varepsilon_{L}$ obeys the volume-dynamics equation
\begin{linenomath}
\begin{equation}\label{eq:vd}
\frac{\partial \varepsilon_{L}}{\partial \tau} + \frac{\partial \varepsilon_{L}}{\partial \vec{L}} \cdot c \cdot \vec{e}_{v} = 0,
\end{equation}
\end{linenomath}
where $\vec{e}_{v}$ is the unit velocity of a portion moving at the speed of light $c$. As shown in Appendix~\ref{app:derivations}, Equation~\eqref{eq:vd} implies the Schrödinger equation, gravity and the curvature of spacetime, and is in precise accordance with observation \cite{Carmesin2021a,Carmesin2026}.

In a homogeneous early universe the locally formed volume (LFV) driven by the dark-energy density $\rho_{vol}$ reproduces the observed value $\Omega_{vol} = \rho_{vol}/\rho_{cr.,0} = 2/3$ at the cosmic microwave background \cite{Planck2020}. Because the indivisible volume portions are quantized, the volume they form does not average to zero: in a heterogeneous universe the vanishing mean field $\langle \vec{G}^{*} \rangle$ is accompanied by a non-zero mean square $\langle (\vec{G}^{*})^{2} \rangle$, which drives additional LFV. As the redshift decreases the heterogeneity grows, this contribution increases, and the resulting heterogeneous Hubble constant increases, reproducing observations at many redshifts and explaining the Hubble tension \cite{Carmesin2026}. For each redshift $z$ it reads
\begin{linenomath}
\begin{equation}\label{eq:H0het}
H_{0,\mathrm{het}}(z) = H_{0,\Lambda\mathrm{CDM}} \cdot \frac{1}{\sqrt{\rho_{cr.,0}}} \cdot \sqrt{\underbrace{\rho_{m}}_{\rho_{m,\mathrm{het}}} + \underbrace{\rho_{vol} \cdot (1 + \kappa(z))^{\xi(z)}}_{\rho_{\Lambda,\mathrm{het}}}},
\end{equation}
\end{linenomath}
with the dynamic density of dark energy $\rho_{vol} = H_{0}^{2}/(4\pi G)$, the auxiliary function $\kappa(z) = \Omega_{m}\,\sigma_{8}/[2\,\Omega_{\mathrm{vol}}\,(1 + z)^{5/2}]$, which quantifies the heterogeneity growing toward low redshift, and the exponent $\xi(z)$ given in Appendix~\ref{app:derivations}. Introducing the critical density of the heterogeneous Universe and the corresponding density parameters (Appendix~\ref{app:derivations}) yields the dark-energy density parameter
\begin{linenomath}
\begin{equation}\label{eq:OLhet}
\Omega_{\Lambda,\mathrm{het}} = \frac{\rho_{\Lambda}}{\rho_{cr.,\mathrm{het}}} = \frac{\rho_{vol} \cdot (1 + \kappa)^{\xi}}{\rho_{m} + \rho_{vol} \cdot (1 + \kappa)^{\xi}} = \frac{\Omega_{vol} \cdot (1 + \kappa)^{\xi}}{\Omega_{m} + \Omega_{vol} \cdot (1 + \kappa)^{\xi}}.
\end{equation}
\end{linenomath}

Equation~\eqref{eq:OLhet} is the sought-after relation for the temporal evolution of the dark-energy density parameter: since $\Omega_{\Lambda,\mathrm{het}}$ depends on $\kappa$ and $\xi$, which in turn depend on the redshift $z$ and thus on cosmic time $t$, it describes how the dark-energy density parameter changes throughout the history of the heterogeneous Universe. Finally, integrating the redshift-dependent expansion history yields the cosmic time evolution and, from the difference between the homogeneous and heterogeneous models, the global time difference $\Delta t_{U}(z)$; this is carried out below.

This behavior admits a physical clarification that explains why the dark-energy density parameter increases toward the present rather than remaining fixed. Heterogeneity causes a variation of gravitational fields $\langle (\vec{G}^{*})^{2} \rangle$. This dynamically generates an additional heterogeneous rate $\langle \dot{\varepsilon}_{L}^{2} \rangle$ of local formation of volume. That rate produces relative additional volume $\langle \varepsilon_{L}^{2} \rangle$ in a heterogeneous manner. This heterogeneous relative additional volume $\langle \varepsilon_{L}^{2} \rangle$ in nature corresponds to dark energy, and $\langle \varepsilon_{L}^{2} \rangle$ substantially increases the dark energy density $\rho_{\Lambda}$ and its density parameter $\Omega_{\Lambda}$. Conversely, in the early Universe, the energy was distributed homogeneously. Therefore, no fields become dynamically active, and no heterogeneous additional volume is formed. Consequently, the dark energy density was at its minimum $\rho_{vol}$.

\noindent\textbf{Cosmological parameters.} The Universe and its time evolution can be described by six cosmological parameters: the Hubble constant $H_{0,\Lambda\mathrm{CDM}}$, or $H_{0,\mathrm{het}}$, the density parameter of curvature $\Omega_{k}$, the density parameter of radiation $\Omega_{r}$, the density parameter of matter $\Omega_{m}$, the density parameter of dark energy $\Omega_{\Lambda}$ or $\Omega_{\mathrm{vol}}$, and the standard deviation of matter fluctuations $\sigma_{8}$. Three of them are derived in the present work, each in a form the reader can follow step by step. First, the density parameter of dark energy follows from the volume dynamics alone: without any specification of how dark energy is composed internally, the formation process of volume yields $\Omega_{\mathrm{vol}} = 2/3$, and Equation~\eqref{eq:OLhet} provides its heterogeneous evolution $\Omega_{\Lambda,\mathrm{het}}(z)$, derived above together with $H_{0,\mathrm{het}}(z)$. Second, the curvature parameter vanishes: observations are consistent with spatial flatness \cite{Planck2020}, and $\Omega_{k} = 0$ has been derived within the present framework \cite{Carmesin2023a}, with two independent proofs given in \cite{Carmesin2023b}. Third, the density parameter of matter follows as the difference $\Omega_{m} = 1 - \Omega_{\Lambda} - \Omega_{r} - \Omega_{k}$.

The remaining parameters are taken from observation. The Hubble constant $H_{0,\Lambda\mathrm{CDM}}$ represents a calendar date $t_{H_{0}} = 1/H_{0}$, the epoch at which we happen to observe the Universe. Therefore, it cannot be derived from first principles; it must be obtained with the help of a measurement. Density parameters, being ratios to the critical density, are free of that calibration, whereas absolute densities such as $\rho_{\Lambda} = \Omega_{\Lambda}\,\rho_{\mathrm{cr.}}$ inherit the epoch of observation through $\rho_{\mathrm{cr.}} = 3H_{0}^{2}/(8\pi G)$; the ratios are therefore what the present derivations provide. The parameter $\sigma_{8}$ is measured. The density parameter of radiation follows from the measured redshift $z_{\mathrm{eq}}$ of matter-radiation equality: since $\rho_{m} \propto (1 + z)^{3}$ and $\rho_{r} \propto (1 + z)^{4}$, equality implies $\Omega_{r} = \Omega_{m}/(1 + z_{\mathrm{eq}})$, which converges after a single iteration started from $\Omega_{m} \approx 1/3$; here $\Omega_{r}$ comprises photons and relativistic neutrinos, as in the determination of $z_{\mathrm{eq}}$ \cite{Planck2020}.

\noindent\textbf{Physics before the classical Big Bang.} As a consequence of gravity and quantum physics, the density in nature is bounded by one half of the Planck density, $\rho_{P}/2$ with $\rho_{P} = 5.155 \times 10^{96}\,\mathrm{kg}/\mathrm{m}^{3}$. At a time $t_{1}$ in the past, the Universe reached $\rho_{1} \approx \rho_{P}/2$, with a present-day light horizon as small as $R_{1} \approx 8.2 \times 10^{-6}\,\mathrm{m}$ \cite{Carmesin2021b}, still far above the Planck length $L_{P} = 1.616 \times 10^{-35}\,\mathrm{m}$. The Big Bang is therefore completed by the physics before the classical Big Bang (BCBB) \cite{Carmesin2021a,Carmesin2021b}: indivisible volume portions exhibit $p = D - 1$ transverse modes and can thus span $D$-dimensional space, so the BCBB started at $D \approx 301$, $R \approx L_{P}$ and $\rho \approx \rho_{P}/2$; expansion and consecutive dimensional phase transitions unfolded the cosmos until the state at $R_{1}$ and $D = 3$ was reached, whereupon the classical Big Bang followed. Both processes have been modeled in a computer simulation \cite{Carmesin2021a,Carmesin2021b}. In particular, the horizon and flatness problems, commonly addressed by cosmic inflation \cite{Guth1981}, are thereby solved from first principles \cite{Carmesin2023a}. Moreover, the states at $D \geq 3$ provide an excitation spectrum of space, from which the formation of neutrinos can be treated.

Altogether, among the six cosmological parameters, $H_{0,\Lambda\mathrm{CDM}}$ represents a calendar date and must therefore be measured. Of the remaining five, three are derived here rather than fitted, and they agree with observation within the stated uncertainties. A derivation of $\sigma_{8}$ and of the radiation content from the physics before the classical Big Bang, based on the excitation spectrum of volume portions and the formation of neutrinos \cite{Carmesin2021a,Carmesin2021b}, is the subject of a forthcoming publication.

\section{Evolution of \texorpdfstring{$\Omega_{\Lambda,\mathrm{het}}(z)$}{Omega\_Lambda,het(z)} and Empirical Validation}\label{sec:validation}

\begin{figure}[!t]
\includegraphics[width=8.5cm]{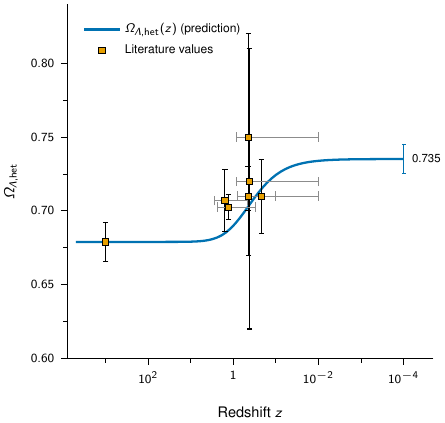}
\caption{Comparison of the theoretical $\Omega_{\Lambda,\mathrm{het}}(z)$ curve with seven literature values. Vertical error bars show the reported uncertainties (statistical and systematic contributions combined in quadrature where reported separately); gray horizontal bars indicate the redshift intervals covered by the underlying data (Table~\ref{tab:OmegaLambda}). Determining $\Omega_{\Lambda}$ requires differences of observables across a redshift interval; at low redshift these differences are necessarily small, which explains the larger uncertainties there and the absence of determinations at $z \ll 1$ (the logarithmic axis expands this region). The supernova analyses of Perlmutter et al. \cite{Perlmutter1999} and Knop et al. \cite{Knop2003} partially share their samples and are therefore not mutually independent. The blue vertical bar indicates the propagated worst-case uncertainty of the predicted present-day value, $\Omega_{\Lambda,\mathrm{het}}(0) = 0.735 \pm 0.010$. The literature values are effective parameters inferred under homogeneous assumptions from the data of the respective redshift range and are shown at the mean probe redshift, in analogy with published redshift-dependent $H_{0}$ analyses; they are not determinations of an instantaneous $\Omega_{\Lambda}(z)$.\label{fig:OmegaLambda}}
\end{figure}

The derived density parameter of dark energy in the heterogeneous Universe,
\begin{linenomath}
\begin{equation*}
\Omega_{\Lambda,\mathrm{het}}(z) = \frac{\Omega_{\mathrm{vol}}\,(1 + \kappa(z))^{\xi(z)}}{\Omega_{m} + \Omega_{\mathrm{vol}}\,(1 + \kappa(z))^{\xi(z)}},
\end{equation*}
\end{linenomath}
with the measured CMB value $\Omega_{\mathrm{vol}} = 0.679$ used in place of the theoretical value $2/3$ (Appendix~\ref{app:derivations}) and $\Omega_{m} = 0.321$ \cite{Planck2020}, increases monotonically as the redshift decreases: from $\Omega_{\Lambda,\mathrm{het}} \approx 0.679$ at high redshift (the homogeneous early Universe) toward $0.735 \pm 0.010$ near $z = 0$. The full evolution is shown in Figure~\ref{fig:OmegaLambda}.

As an empirical test, the predicted curve is compared in Figure~\ref{fig:OmegaLambda} with literature values that span several orders of magnitude in redshift and originate from mutually independent methods (CMB analyses, baryon acoustic oscillations, gravitational-wave catalogs, type Ia supernova compilations, X-ray cluster abundances and galaxy/quasar observations). In particular, the low-redshift value $\Omega_{\Lambda,\mathrm{het}}(z = 0) = 0.735 \pm 0.010$ agrees with independent determinations \cite{Riess2000,Perlmutter1998}, while the high-precision Planck value ($z \approx 1000$, $\Omega_{\Lambda} = 0.679 \pm 0.013$) \cite{Planck2020} fixes the high-redshift end of the curve rather than testing it, as discussed below. The remaining points likewise lie close to the prediction, their deviations being consistent with the differing methods and uncertainties; the complete list of literature values is given in Table~\ref{tab:OmegaLambda}. This broad agreement across the observed redshift range supports the physical relevance of the heterogeneous dark-energy model.

\begin{figure}[!t]
\includegraphics[width=8.5cm]{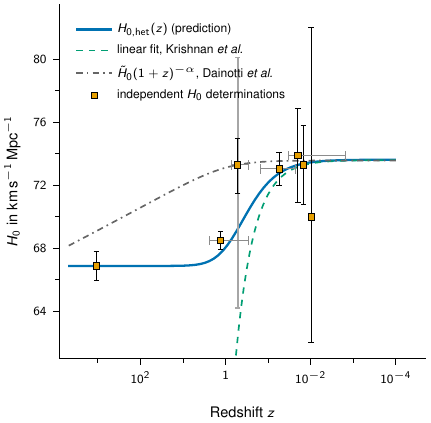}
\caption{Comparison of the theoretical $H_{0,\mathrm{het}}(z)$ curve, Equations~\eqref{eq:E32}--\eqref{eq:E34}, with seven independent determinations of $H_{0}$ from early- and late-Universe probes. Error bars and gray horizontal redshift intervals as in Figure~\ref{fig:OmegaLambda}, with redshifts of extended samples represented by their means (Table~\ref{tab:H0}). The nominally high strong-lensing value becomes fully consistent with the prediction under the enlarged uncertainty of the mass-profile-independent reanalysis of the same lens systems. The corresponding comparison within the volume dynamics was first presented in \cite{Carmesin2026}. The lighter, wider bar on the strong-lensing point spans 64.2--80.1 km\,s$^{-1}$\,Mpc$^{-1}$, the interval implied by the mass-profile-independent TDCOSMO reanalyses of the same lens systems \cite{Birrer2020}; within it, the determination is consistent with the prediction. As in Figure~\ref{fig:OmegaLambda}, the plotted values are effective parameters inferred under homogeneous assumptions from the data of the respective redshift range; they are not determinations of an instantaneous quantity. The dashed line and the dash-dotted curve are the functions fitted to redshift-binned determinations of $H_{0}$ by Krishnan \textit{et al.} \cite{Krishnan2020} and Dainotti \textit{et al.} \cite{Dainotti2021}, discussed in Section~\ref{sec:binned}. The fitted line leaves the panel at its lower edge and reaches $H_{0} = 0$ at $z \approx 3.4$.\label{fig:H0}}
\end{figure}

The literature values entering this comparison require a remark on their interpretation. Each of them is an effective parameter, inferred from the data of a given redshift range under the assumption of a homogeneous background, and each is shown at the mean redshift of the underlying probe. They are therefore not determinations of an instantaneous $\Omega_{\Lambda}(z)$, and Figure~\ref{fig:OmegaLambda} is not a point-by-point measurement of the derived function. The same reading underlies published analyses in which the Hubble constant is inferred separately in redshift bins and a trend with redshift is reported \cite{Krishnan2020,Dainotti2021}: within each bin the parameter is treated as constant, while the collection of bins is examined for evolution. If the parameters are in fact redshift dependent, an analysis of this kind returns an average over the interval it covers rather than a local value, so that the effective values are expected to follow the derived curve without coinciding with it exactly. The gray horizontal bars in Figures~\ref{fig:OmegaLambda} and \ref{fig:H0} indicate the intervals over which each average is taken. These two analyses, and the functions they fit to their binned values, are examined in Section~\ref{sec:binned}.

\begin{table}[!t]
\caption{The following literature values of the dark-energy density parameter $\Omega_{\Lambda}$ at various redshifts $z$ are compared with the heterogeneous prediction $\Omega_{\Lambda,\mathrm{het}}(z)$ in Figure~\ref{fig:OmegaLambda}; the entry marked with $^{\dagger}$ is listed for completeness only and is not included in Figure~\ref{fig:OmegaLambda} (see the note below the table). They originate from mutually independent methods (CMB analyses, baryon acoustic oscillations, gravitational-wave transient catalogs, type Ia supernova compilations, X-ray cluster abundances and galaxy/quasar observations); dependences between individual determinations within the supernova category are noted below. Where a redshift interval is reported in the source, its mean is given. The recovery of one and the same predicted curve $\Omega_{\Lambda,\mathrm{het}}(z)$ by these very different observational techniques considerably strengthens the empirical support for the predicted time evolution: the behavior is not tied to any particular method but appears consistently across all of them, as expected for a fundamental property of the Universe rather than for a systematic effect of an individual probe. As the comparison combines electromagnetic observations, ranging from the CMB to supernovae, quasars and X-ray clusters, with gravitational-wave data, it follows the approach of multi-messenger astronomy.\label{tab:OmegaLambda}}
\begin{adjustwidth}{-\extralength}{0cm}
\small
\begin{tabularx}{\fulllength}{lllL}
\toprule
\textbf{Reference} & \textbf{Redshift} \boldmath{$z$} & \boldmath{$\Omega_{\Lambda}$} & \textbf{Method}\\
\midrule
Planck Collaboration (2018/2020) \cite{Planck2020} & $\approx 1000$ & $0.679 \pm 0.013$ & CMB temperature/polarization\\
Escamilla-Rivera \& Nájera (2022) \cite{Escamilla2022} & $\approx 0.22$ & $0.71 \pm 0.025$ & Gravitational-wave transient catalogs\\
DESI Collaboration (2025) \cite{DESI2025} & $\approx 1.31$ & $0.7025 \pm 0.0086$ & Baryon acoustic oscillations\\
Cao et al. (2020) \cite{Cao2020} & $\approx 1.58$ & $0.707 \pm 0.021$ & H\,\textsc{ii} galaxies, quasar angular size\\
Perlmutter et al. (1999) \cite{Perlmutter1999} & $\approx 0.42$ & $0.72^{+0.09}_{-0.10}$ & Type Ia supernovae\\
Knop et al. (2003) \cite{Knop2003} & $\approx 0.44$ & $0.75^{+0.07}_{-0.08}$ & Type Ia supernovae\\
Brout et al. (2022) \cite{Brout2022} $^{\dagger}$ & $\approx 1.13$ & $0.625 \pm 0.084$ & Type Ia supernovae (Pantheon+)\\
Ghirardini et al. (2024) \cite{Ghirardini2024} & $\approx 0.45$ & $0.71^{+0.02}_{-0.01}$ & X-ray cluster abundance\\
\bottomrule
\end{tabularx}
\end{adjustwidth}
\noindent{\footnotesize For Escamilla-Rivera \& Nájera, the redshift intervals $[0.01; 0.49]$ and $[0.03; 0.8]$ reported in their Fig.~7 were averaged to $z \approx 0.22$. For Cao et al., the matter density $\Omega_{m,0} = 0.293 \pm 0.021$ is reported, from which $\Omega_{\Lambda} \approx 0.707 \pm 0.021$ follows for a spatially flat universe, and the redshift interval $z \in [0.46; 2.7]$ was averaged to $z \approx 1.58$. For the DESI Collaboration, the matter density $\Omega_{m} = 0.2975 \pm 0.0086$ is reported for the spatially flat $\Lambda$CDM model from baryon acoustic oscillation data alone, from which $\Omega_{\Lambda} = 0.7025 \pm 0.0086$ follows; the redshift interval $z \in [0.295; 2.33]$ covered by the DESI tracers was averaged to $z \approx 1.31$. For Perlmutter et al., the matter density $\Omega_{M} = 0.28^{+0.09}_{-0.08}\,\text{(statistical)}\,^{+0.05}_{-0.04}\,\text{(systematic)}$ is reported for a spatially flat universe, from which $\Omega_{\Lambda} = 0.72^{+0.08}_{-0.09}\,\text{(statistical)}\,^{+0.04}_{-0.05}\,\text{(systematic)}$ follows; the statistical and systematic uncertainties were combined in quadrature and the redshift interval $z \in [0.01; 0.83]$ of the underlying supernovae was averaged to $z \approx 0.42$. For Knop et al., $\Omega_{\Lambda} = 0.75^{+0.06}_{-0.07}\,\text{(statistical)} \pm 0.04\,\text{(systematic)}$ is reported for a spatially flat universe; the uncertainties were combined in quadrature and the redshift interval $z \in [0.01; 0.86]$ was averaged to $z \approx 0.44$. As the analysis of Knop et al. combines its new supernovae with the earlier sample of Perlmutter et al., these two determinations are not statistically independent of each other. For Ghirardini et al., the matter density $\Omega_{m} = 0.29^{+0.01}_{-0.02}$ is reported for the spatially flat $\Lambda$CDM model from the abundance of X-ray selected galaxy clusters in the first SRG/eROSITA All-Sky Survey with weak-lensing mass calibration, from which $\Omega_{\Lambda} = 0.71^{+0.02}_{-0.01}$ follows; the redshift interval $z \in [0.1; 0.8]$ of the cluster sample was averaged to $z \approx 0.45$.\par}

\noindent{\footnotesize $^{\dagger}$For the Pantheon+ compilation of Brout et al., the tabulated value $\Omega_{\Lambda} = 0.625 \pm 0.084$ is the determination obtained without the assumption of spatial flatness, and the redshift interval $z \in [0.001; 2.26]$ of the 1550 supernovae was averaged to $z \approx 1.13$. This value is listed for completeness but is not included in Figure~\ref{fig:OmegaLambda}: spatial flatness has been demonstrated both experimentally \cite{Planck2020} and theoretically \cite{Carmesin2023a}, so a determination that does not presuppose flatness is not based on adequate premises. Under the assumption of spatial flatness, the reported matter density $\Omega_{M} = 0.334 \pm 0.018$ corresponds to $\Omega_{\Lambda} = 0.666 \pm 0.018$. Even under this assumption, the corresponding matter density lies about $2\sigma$ above the independent determinations from baryon acoustic oscillations \cite{DESI2025} and cluster abundances \cite{Ghirardini2024}; this known tension between modern type Ia supernova compilations and other cosmological probes contributes, in combined analyses, to the reported evidence for a time-evolving dark-energy density \cite{DESI2025}. A direct confrontation of the prediction $\Omega_{\Lambda,\mathrm{het}}(z)$ with the supernova Hubble diagram itself, rather than with parameters fitted under the premise of a dark-energy density that is constant across the entire redshift range of the compilation, is left for future work.\par}
\end{table}

A further qualification concerns the high-redshift end of both curves. There the heterogeneity vanishes, $\kappa(z) \to 0$, so that $\Omega_{\Lambda,\mathrm{het}} \to \Omega_{\mathrm{vol}}$ and $H_{0,\mathrm{het}} \to H_{0,\Lambda\mathrm{CDM}}$, and the curves approach the measured input values, $\Omega_{\mathrm{vol}} = 0.679$ in Figure~\ref{fig:OmegaLambda} and $H_{0,\Lambda\mathrm{CDM}} = 66.88$\,km\,s$^{-1}$\,Mpc$^{-1}$ in Figure~\ref{fig:H0}. The Planck points therefore anchor the curves rather than test them, and what the comparison probes is the evolution toward low redshift. The test at high redshift lies elsewhere. The volume dynamics yields $\Omega_{\mathrm{vol}} = 2/3$ without adjustment to any datum, and this derived value agrees with the measured $\Omega_{\Lambda}(z_{CMB}) = 0.679 \pm 0.013$ \cite{Planck2020} within one standard deviation; the same agreement in terms of the energy densities is shown in Appendix~\ref{app:derivations}.

That derivation belongs to the first of three stages of modeling, and keeping the stages apart prevents a misunderstanding that the notation invites. (i) Volume formation is analyzed in empty space and yields the dark-energy density $\rho_{\mathrm{vol}}$, hence $\Omega_{\mathrm{vol}} = 2/3$. Empty space contains nothing that could be diluted by the expansion, so no density evolves in time at this stage, and the Hubble time $t_{H_{0}}$ is defined while an age of the Universe is not. (ii) Homogeneous matter and radiation are then placed in the formed space, which gives the $\Lambda$CDM model; only here does the integrated expansion history yield an age, because $\Omega_{m}$ and $\Omega_{r}$ enter the integrand. (iii) Heterogeneity is added last, as developed in this work. Stages (ii) and (iii) express their results as ratios with respect to the stage before, $\Omega_{\Lambda,\mathrm{het}}/\Omega_{\mathrm{vol}}$, $H_{0,\mathrm{het}}(z)/H_{0,\Lambda\mathrm{CDM}}$ and $t_{U,\mathrm{het}}/t_{U,\mathrm{hom}}$, so that the result of the preceding stage is used but never reopened. Reading the relation $\rho_{\Lambda} = \Omega_{\Lambda}\,\rho_{\mathrm{cr.}}$ across the stages would suggest that $\rho_{\mathrm{vol}}$ falls as the critical density falls. It does not: within stage (i) there is no time evolution of any density, and the later stages modify that result by the stated ratios rather than recomputing it. The relation from which $\rho_{\mathrm{vol}}$ follows, Equation~\eqref{eq:E19}, connects the rate of volume formation with $\rho_{\mathrm{vol}}$ and the Hubble time alone, and dividing by the critical density removes the epoch of observation and leaves the pure number $2/3$, Equation~\eqref{eq:E22}. The epoch dependence of absolute densities noted in Section~\ref{sec:derivation} therefore concerns their calibration against a measured $H_{0}$, not an evolution within stage (i).

A limitation of this test should be stated openly. The comparison confronts the derived functions with cosmological parameters as published by the respective analyses, not with the underlying data sets themselves. No likelihood analysis of supernova magnitudes, baryon acoustic oscillation distances or cluster counts against the heterogeneous model has been carried out, so the agreement seen in Figures~\ref{fig:OmegaLambda} and \ref{fig:H0} is not a goodness-of-fit statement. Two considerations motivate this choice. First, apart from the measured inputs that fix the high-redshift end, the derived functions contain no parameter that is adjusted to the data entering the comparison, so what is shown is a test of a prediction and not a fit; in particular, the evolution toward low redshift follows from the derivation and is not tuned to the values it is compared with. Second, a joint reanalysis of the underlying catalogs would require the full covariance structure of each sample together with a consistent treatment of the calibration and nuisance parameters of each method, which lies beyond the scope of this work. Such a direct confrontation, in particular with the supernova Hubble diagram itself, is left for future work.

An analogous comparison holds for the expansion rate itself, see Figure~\ref{fig:H0}: the derived $H_{0,\mathrm{het}}(z)$ rises from the Planck value at high redshift toward the locally measured range, in agreement with independent early- and late-Universe determinations of $H_{0}$ (Table~\ref{tab:H0}).

\begin{table}[!t]
\caption{The following independent determinations of the Hubble constant are compared with the derived $H_{0,\mathrm{het}}(z)$ in Figure~\ref{fig:H0}. Redshifts of extended samples are represented by their means; all values of $H_{0}$ are given in km\,s$^{-1}$\,Mpc$^{-1}$.\label{tab:H0}}
\begin{adjustwidth}{-\extralength}{0cm}
\small
\begin{tabularx}{\fulllength}{lllL}
\toprule
\textbf{Reference} & \textbf{Redshift} \boldmath{$z$} & \boldmath{$H_{0}$} & \textbf{Method}\\
\midrule
Planck Collaboration (2018/2020) \cite{Planck2020} & $\approx 1090$ & $66.88 \pm 0.92$ & CMB temperature/polarization\\
DESI Collaboration (2025) \cite{DESI2025} & $\approx 1.31$ & $68.51 \pm 0.58$ & Baryon acoustic oscillations with BBN prior\\
Wong et al. (2020) \cite{Wong2020} & $\approx 0.5$ & $73.3^{+1.7}_{-1.8}$ & Strong-lensing time delays (H0LiCOW)\\
Riess et al. (2022) \cite{Riess2022} & $\approx 0.055$ & $73.04 \pm 1.04$ & Type Ia supernovae, Cepheid distance ladder\\
Pesce et al. (2020) \cite{Pesce2020} & $\approx 0.02$ & $73.9 \pm 3.0$ & Water megamasers\\
Blakeslee et al. (2021) \cite{Blakeslee2021} & $\approx 0.015$ & $73.3 \pm 2.5$ & Surface brightness fluctuations\\
Abbott et al. (2017) \cite{Abbott2017} & $\approx 0.01$ & $70^{+12}_{-8}$ & Gravitational-wave standard siren GW170817\\
\bottomrule
\end{tabularx}
\end{adjustwidth}
\noindent{\footnotesize For Wong et al., $z \approx 0.5$ is the mean deflector redshift of the six lensed systems. The reported 2.4 percent uncertainty of this determination rests on specific assumptions about the deflector mass profiles; the hierarchical reanalysis of the same systems within TDCOSMO, in which the mass-sheet degeneracy is constrained by stellar kinematics alone, broadens the constraint to $H_{0} = 74.5^{+5.6}_{-6.1}$ km\,s$^{-1}$\,Mpc$^{-1}$ and shifts it to $H_{0} = 67.4^{+4.1}_{-3.2}$ km\,s$^{-1}$\,Mpc$^{-1}$ when external lens samples are added \cite{Birrer2020}. Within these uncertainties, the strong-lensing determination is fully consistent with the predicted $H_{0,\mathrm{het}}(z)$. Moreover, for time-delay distances, which depend on both the deflector and the source redshifts, the assignment of a single emission redshift is less direct than for the other probes. For Riess et al., $z \approx 0.055$ is the mean redshift of the calibrating supernova sample. For Pesce et al. and Blakeslee et al., the redshifts represent the means of the megamaser-host and galaxy samples, respectively. For Blakeslee et al., the statistical and systematic uncertainties $\pm 0.7$ and $\pm 2.4$ were combined in quadrature. For Abbott et al., the redshift of the host galaxy NGC~4993 is used. The gray horizontal bars in Figure~\ref{fig:H0} span the documented sample ranges: $z \in [0.295; 2.33]$ for the DESI BAO tracers, $z \in [0.295; 0.745]$ for the H0LiCOW deflectors, $z \in [0.023; 0.15]$ for the SH0ES Hubble-flow supernovae and $z \in [0.0015; 0.034]$ for the megamaser hosts (NGC~4258 to NGC~6264). For Blakeslee et al., the sample extends to 100~Mpc, with the majority of galaxies between 50 and 80~Mpc.\par}
\end{table}

\section{Derivation of the Time Evolution of Global Time Dilation}\label{sec:time}

\subsection{Motivation and Approach}\label{sec:motivation}

While the standard $\Lambda$CDM framework assumes a homogeneous universe with constant density parameters, the results above show that this is an idealization: in a realistic, heterogeneous universe the Hubble parameter $H$, the Hubble constant $H_{0}$ and the dark-energy density parameter $\Omega_{\Lambda}$ evolve with redshift because of structure formation, including the formation of heterogeneity. The temporal evolution of the Universe must therefore be reconsidered, and the age is derived below as a function of redshift, allowing a direct comparison between the homogeneous and the heterogeneous model.

The age $t_{U}(z)$ of the Universe is obtained, for both the homogeneous and the heterogeneous case, by integrating the relation between the scale factor and cosmic time, with the redshift-dependent $H_{0,\mathrm{het}}(z)$ and $\Omega_{\Lambda,\mathrm{het}}(z)$ derived above entering the integrand. The full integral, the auxiliary functions $\kappa(z)$ and $\xi(z)$, and the numerical scheme are given in Appendix~\ref{app:derivations}.

\subsection{Age Difference and Time Dilation}\label{sec:agediff}

Evaluating the integrated expansion history numerically for the present day, $x_{2} = 1 =: x_{0}$, gives the present-day age of the heterogeneous Universe
\begin{linenomath}
\begin{equation}\label{eq:tUhet}
t_{U,\mathrm{het}}(x_{0}) = 13.521^{+0.355}_{-0.338}\,\mathrm{Gyr}.
\end{equation}
\end{linenomath}
The corresponding age of the homogeneous Universe, obtained by replacing $\Omega_{\Lambda,\mathrm{het}}$ with $\Omega_{\mathrm{vol}} = \Omega_{\Lambda,\mathrm{hom}}$ and $\Omega_{m,\mathrm{het}}$ with $\Omega_{m}$, is
\begin{linenomath}
\begin{equation}\label{eq:tUhom}
t_{U,\mathrm{hom}}(x_{0}) = 13.827^{+0.357}_{-0.340}\,\mathrm{Gyr}.
\end{equation}
\end{linenomath}
The time difference $\Delta t_{U} = t_{U,\mathrm{hom}} - t_{U,\mathrm{het}}$, Equation~\eqref{eq:E38}, represents a global time dilation of the homogeneous Universe: between the same two events, the homogeneous Universe exhibits a larger time interval than the heterogeneous Universe.

\begin{figure}[H]
\includegraphics[width=8.5cm]{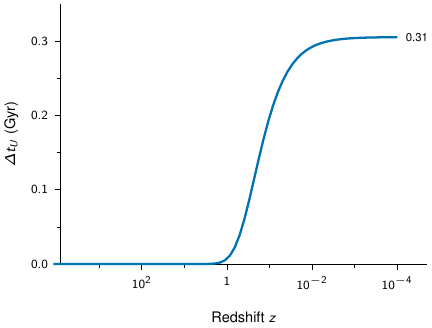}
\caption{The global time difference $\Delta t_{U} = t_{U,\mathrm{hom}} - t_{U,\mathrm{het}}$ as a function of the redshift $z$.\label{fig:DeltatUz}}
\end{figure}

In Figure~\ref{fig:DeltatUz}, that global time difference $\Delta t_{U} = t_{U,\mathrm{hom}} - t_{U,\mathrm{het}}$ is shown as a function of the redshift $z$; at the present day it amounts to $\Delta t_{U} = 0.306 \pm 0.013$\,Gyr. An alternative representation as a function of the scaled scale radius $x$ is given in Figure~\ref{fig:DeltatUx}.

\section{Comparison with Redshift-Binned Analyses of \texorpdfstring{$H_{0}$}{H0}}\label{sec:binned}

That the Hubble constant inferred from observations is not constant across redshift was reported before the present work. Krishnan \textit{et al.} \cite{Krishnan2020} found a descending trend in a combined low-redshift data set, and Dainotti \textit{et al.} \cite{Dainotti2021} found the same behavior in the Pantheon supernova compilation. Both results are empirical: the values obtained in the individual bins are fitted with a function chosen for convenience, and the trend is read off from its parameters. The present work builds on these findings. What those analyses established for a limited range of redshift follows here from a derivation, and it extends to the full range up to the microwave background. The two fitted functions are shown together with the derived $H_{0,\mathrm{het}}(z)$ in Figure~\ref{fig:H0}.

\subsection{The Descending Line of Krishnan et al.}\label{sec:krishnan}

Krishnan \textit{et al.} \cite{Krishnan2020} bin megamasers, cosmic chronometers, type Ia supernovae and baryon acoustic oscillations at $z \leq 0.7$ into six bins and fit the flat $\Lambda$CDM model separately in each. The resulting values of $H_{0}$ descend with redshift, and a weighted straight-line fit gives a slope of $-21.7 \pm 9.4$\,km\,s$^{-1}$\,Mpc$^{-1}$ per unit redshift with an intercept of $73.6 \pm 2.5$\,km\,s$^{-1}$\,Mpc$^{-1}$, the slope differing from a horizontal line at $2.1\sigma$. The intercept is an outcome of that fit rather than an input, and it agrees with the present-day value of the derived function, $H_{0,\mathrm{het}}(0) = 73.6$\,km\,s$^{-1}$\,Mpc$^{-1}$. Over the fitted interval the derived function descends more slowly: its mean slope between $z = 0$ and $z = 0.61$ is $-7.7$\,km\,s$^{-1}$\,Mpc$^{-1}$ per unit redshift, which lies 1.5 standard deviations from the fitted slope.

A straight line cannot, however, describe the behavior outside the interval in which it was obtained. Continued to higher redshift it reaches $H_{0} = 0$ at $z \approx 3.4$ and turns negative beyond. As the zero is approached the Hubble time $t_{H} = 1/H_{0}$ diverges, so that the implied age of the Universe grows without bound, which the cosmic-clock constraint of Tomasetti \textit{et al.} \cite{Tomasetti2026} excludes. The line is therefore a local approximation to the measured trend, valid where it was fitted, and the feature it identifies calls for a function that also possesses a high-redshift limit. The derived $H_{0,\mathrm{het}}(z)$ provides one: as the heterogeneity vanishes, $\kappa(z) \to 0$, the function approaches the value measured at the microwave background.

\subsection{The Evolutionary Parameterization of Dainotti et al.}\label{sec:dainotti}

Dainotti \textit{et al.} \cite{Dainotti2021} divide the Pantheon compilation of 1048 supernovae into 3, 4, 20 and 40 redshift bins and fit the extracted values with $H_{0}(z) = \tilde{H}_{0}(1 + z)^{-\alpha}$. For three bins in the flat $\Lambda$CDM model they obtain $\tilde{H}_{0} = 73.577 \pm 0.106$\,km\,s$^{-1}$\,Mpc$^{-1}$ and $\alpha = 0.009 \pm 0.004$, the evolutionary parameter differing from zero at $2.0\sigma$; the other binnings give values consistent with these. Here $\tilde{H}_{0}$ is not an independent determination, because the absolute magnitude of the supernovae is calibrated so that $H_{0} = 73.5$\,km\,s$^{-1}$\,Mpc$^{-1}$ at $z = 0$. The informative quantity is the exponent, and it confirms the descending trend.

With an exponent of order $10^{-2}$ the function falls very slowly. Extrapolated to the last scattering surface it yields $69.2 \pm 2.2$\,km\,s$^{-1}$\,Mpc$^{-1}$, which remains above the measured $66.88 \pm 0.92$\,km\,s$^{-1}$\,Mpc$^{-1}$ \cite{Planck2020}; the compatibility reported by the authors rests on the large uncertainty propagated from $\alpha$. The shape differs as well, as Figure~\ref{fig:H0} shows. A power law of this kind carries no characteristic redshift and therefore stays close to its present-day value throughout, whereas the determinations of Table~\ref{tab:H0} and the derived function fall from the local range toward the microwave-background value between $z \approx 0.01$ and $z \approx 3$. The derived function reproduces that behavior because the heterogeneity on which it rests grows as $\kappa(z) \propto (1 + z)^{-5/2}$, Equation~\eqref{eq:E33}, which fixes both where the transition occurs and where the function levels off.

\subsection{Derivation and Parameterization}\label{sec:parameterization}

Both functions were introduced to test whether the binned values are constant, and for that purpose the choice of function is secondary. Neither is derived, and neither relates the trend to a physical cause: a straight line and a power law contain no quantity that links $H_{0}(z)$ to the structure of the Universe. In the present framework the redshift dependence is not assumed. It follows from the law of locally formed volume, Equation~\eqref{eq:E16}, with the growth of heterogeneity entering through $\sigma_{8}$, so that the shape of the function and its high-redshift limit are fixed before any comparison with data is made. This is what allows a single function to hold from the microwave background to the present day.

\section{Conclusions}\label{sec:conclusions}

Recent observational studies aim to constrain the age of the Universe independently of a single cosmological model. A relevant example is provided by Tomasetti et al. \cite{Tomasetti2026}, who use the ages of the oldest Milky Way stars as cosmic clocks to obtain a lower limit on the age of the Universe, $t_{U} \geq 13.8 \pm 1.0\,\text{(stat.)} \pm 1.4\,\text{(syst.)}$\,Gyr. Such constraints are usually interpreted within a homogeneous $\Lambda$CDM framework, which does not account for the dynamical influence of large-scale heterogeneity on cosmic time. The present work shows that, once heterogeneity is included as a dynamical component, the Hubble parameter, the Hubble constant $H_{0}(z)$ and the dark-energy density parameter $\Omega_{\Lambda}(z)$ become coupled to the time evolution of the Universe. The cosmic age is therefore no longer fixed by the homogeneous expansion history alone, but depends on the model used to describe the real, structured Universe.

Integrating the redshift-dependent expansion history, Equation~\eqref{eq:E31}, yields the self-consistent age of the heterogeneous Universe, $t_{U} = 13.521^{+0.355}_{-0.338}$\,Gyr, which remains compatible with the cosmic-clock bound, lying within one standard deviation of its central value. This indicates that the cosmic-clock age constraint, the observed discrepancy in $H_{0}$ \cite{Riess2022} and the time evolution of $\Omega_{\Lambda}(z)$ need not be regarded as separate problems, but may be different observational manifestations of the same underlying fact: the Universe is not perfectly homogeneous, and its heterogeneity influences the progression of cosmic time. Together with its agreement with independent empirical constraints, the temporal evolution of dark energy derived here forms a coherent framework linking cosmic structure, dark energy and the age of the Universe. Independent late-Universe analyses point in the same direction: baryon acoustic oscillation measurements of the DESI collaboration favor a time-evolving dark energy when combined with supernova and CMB data \cite{DESI2025}, and the Hubble diagram of quasars at high redshift suggests a dark-energy density that increases with cosmic time \cite{Risaliti2019} (see also Figure~\ref{fig:OmegaLambda}). A more detailed discussion, including the interpretation of $t_{U}$, $H_{0}$ and $\sigma_{8}$ as effective averaged quantities and an outlook on future tests, is given in Appendix~\ref{app:effective}.

More broadly, the heterogeneous treatment developed here forms part of a first-principles cosmological framework based on the volume dynamics of space \cite{Carmesin2021a,Carmesin2026} (Figure~\ref{fig:map}), within which the coupling of cosmic structure, dark energy and cosmic time examined in this work arises naturally.

\vspace{6pt}

\authorcontributions{H.-O.C. developed the volume-dynamics framework and the derivation of the heterogeneous Hubble constant $H_{0,\mathrm{het}}(z)$ on which this work builds. J.D.Y. and H.-O.C. jointly derived the temporal evolution of the dark-energy density parameter $\Omega_{\Lambda,\mathrm{het}}(z)$ and the integrated cosmic time evolution. J.D.Y. performed the numerical integration of the expansion history, carried out the comparison with the independent literature values, prepared the figures and wrote the manuscript. H.-O.C. supervised the work. Both authors discussed the results and reviewed the final manuscript.}

\funding{This research received no external funding.}

\dataavailability{All data supporting the findings of this study are available within the article and its appendices. No new observational data were generated or analyzed in this study. The observational values used for the empirical validation are taken from the published literature cited in the text and are compiled in Tables~\ref{tab:OmegaLambda} and \ref{tab:H0}.}

\backnotes{Code Availability}{The numerical integrations of Equations~\eqref{eq:E35} and \eqref{eq:E37}, together with the worst-case uncertainty propagation described in Appendix~\ref{app:derivations}, were performed in two spreadsheet workbooks (Microsoft Excel) that implement the sums given in Appendix~\ref{app:derivations} and cross-reference Equations~\eqref{eq:E36} and \eqref{eq:E38}, one using the Planck parameters and one using the derived parameters. The workbooks are openly available on Zenodo at \url{https://doi.org/10.5281/zenodo.21688383}.}

\conflictsofinterest{The authors declare no competing interests.}

\appendixtitles{yes}
\appendixstart
\appendix
\section[\appendixname~\thesection]{Map of the Derivation}\label{app:map}

\begin{figure}[H]
\begin{adjustwidth}{-\extralength}{0cm}
\centering
\includegraphics[width=\fulllength]{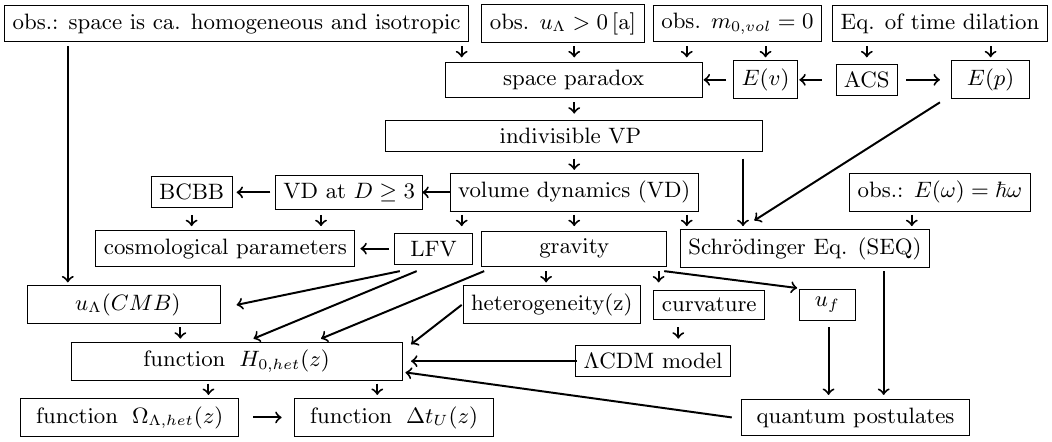}
\end{adjustwidth}
\caption{\underline{Paths of derivation}: each arrow points from a premise to an implied statement or quantity. Very reliable observations (obs.) and the equation of time dilation are the starting points of the deductions. The steps leading to $H_{0}(z)$ are also presented in \cite{Carmesin2026}. The items below the VD, the ACS, $E(v)$, $E(p)$ and $E(\omega)$ have already been confirmed by observation. The VD implies gravity, causing heterogeneity and curvature, which yields the $\Lambda$CDM model. It describes the early (homogeneous) Universe, from which heterogeneity and LFV shape $H_{0,het}$, $\Omega_{\Lambda,het}$ and $\Delta t_{U}$. The quantity $u_{f}$ denotes the energy density of the gravitational field. Abbreviations: VP, volume portion; ACS, adequate coordinate system; BCBB, before classical Big Bang; LFV, locally formed volume; $m_{0,vol}$, rest mass of a VP. [a] The fact $u_{\Lambda} > 0$ is observed directly, or it is founded independently as follows: two masses in empty space exhibit gravity; consequently, momentum is transmitted between these masses. As no other entity is available in empty space, the empty space transmits that momentum; therefore, empty space includes momentum and energy. Consequently, $u_{\Lambda} > 0$. The VD implies the value $\Omega_{\mathrm{vol}} = 2/3 > 50\,\%$; therefore, the energy of space, or of volume, is the energetically dominant component of the Universe.\label{fig:map}}
\end{figure}

The full chain of reasoning, leading from a small set of very reliable observations to the functions $H_{0}(z)$, $\Omega_{\Lambda,\mathrm{het}}(z)$ and $\Delta t_{U}(z)$, is summarized in Figure~\ref{fig:map}.

\section[\appendixname~\thesection]{Homogeneous \texorpdfstring{$\Lambda$}{Lambda}CDM Background and Linear Growth}\label{app:lcdm}

The homogeneous $\Lambda$CDM model provides the reference against which the heterogeneous results of Sections~\ref{sec:homogeneous}--\ref{sec:time} are compared. Dark energy is modeled as a cosmological constant $\Lambda$ with constant density $\rho_{\Lambda} = \Lambda c^{2}/(8\pi G)$, entering the first Friedmann--Lemaître equation, Equation~\eqref{eq:friedmann}. For a spatially flat universe ($k = 0$) the homogeneous Hubble parameter is, in the early Universe ($z > 10^{3}$),
\begin{linenomath}
\begin{equation}\label{eq:B1}
H_{\mathrm{hom}}(z) = H_{0}\,\sqrt{\Omega_{m,0}(1 + z)^{3} + \Omega_{r}(1 + z)^{4} + \Omega_{\Lambda}},
\end{equation}
\end{linenomath}
where the radiation term $\Omega_{r}(1 + z)^{4}$ is relevant only for $z > 10^{3}$, and at lower redshift,
\begin{linenomath}
\begin{equation}\label{eq:B2}
H_{\mathrm{hom}}(z) = H_{0,\Lambda\mathrm{CDM}}\,\sqrt{\Omega_{m,0}(1 + z)^{3} + \Omega_{\Lambda}}.
\end{equation}
\end{linenomath}
With the Planck parameters \cite{Planck2020} $H_{0} \approx 67.66$\,km\,s$^{-1}$\,Mpc$^{-1}$, $\Omega_{m,0} \approx 0.3111$, $\Omega_{\Lambda} \approx 0.6889$ and $\Omega_{r,0} \approx 0.0001$, Equations~\eqref{eq:B1} and \eqref{eq:B2} give, for example, $H_{\mathrm{hom}}(10^{3}) \approx 1.37 \times 10^{6}$, $H_{\mathrm{hom}}(1) \approx 120.6$ and $H_{\mathrm{hom}}(10^{-4}) \approx 67.67$\,km\,s$^{-1}$\,Mpc$^{-1}$ $\approx H_{0}$.

The present-day homogeneous age follows from the Hubble time $t_{H} = 1/H_{0}$. Numerically, $t_{H} = 1/H_{0} \approx 4.57 \times 10^{17}\,\mathrm{s} \approx 14.5$\,Gyr; applying the correction factor 0.95 \cite{Carmesin2024} yields
\begin{linenomath}
\begin{equation}\label{eq:B3}
t_{0,\mathrm{hom}} = t_{H} \times 0.95 = 14.5\,\mathrm{Gyr} \times 0.95 = 13.775\,\mathrm{Gyr},
\end{equation}
\end{linenomath}
in agreement with a typical $\Lambda$CDM value. This is the value quoted in Section~\ref{sec:homogeneous}; the self-consistent homogeneous age obtained by integrating the full expansion history (Equation~\eqref{eq:tUhom}) is $t_{U,\mathrm{hom}} = 13.827^{+0.357}_{-0.340}$\,Gyr.

Heterogeneity is quantified by the parameter $\sigma_{8}$, which enters the auxiliary function $\kappa(z)$, Equation~\eqref{eq:E33}, through the linear growth of structure. The complete linear-growth treatment underlying $\kappa(z)$ and the exponent $\xi(z)$, including the higher-order terms $q_{t}^{(n)}$ of which the first order $q_{1}^{(1)}$ is used in Section~\ref{sec:homogeneous}, is given in \cite{Carmesin2026} (Eqs.~(223), (244) and (237--240) therein) and in \cite{Carmesin2023b,Carmesin2022}.

\section[\appendixname~\thesection]{Strict Monotonicity of \texorpdfstring{$\Omega_{\Lambda,\mathrm{het}}(z)$}{Omega\_Lambda,het(z)}}\label{app:monotonicity}

The dark-energy density parameter of the heterogeneous Universe, Equation~\eqref{eq:OLhet}, depends on redshift only through the single combination $\varphi(z) = [1 + \kappa(z)]^{\xi(z)}$, so that
\begin{linenomath}
\begin{equation}\label{eq:C1}
\Omega_{\Lambda,\mathrm{het}} = \frac{\Omega_{\mathrm{vol}}\,\varphi}{\Omega_{m} + \Omega_{\mathrm{vol}}\,\varphi}.
\end{equation}
\end{linenomath}
The auxiliary function $\kappa(z)$, Equation~\eqref{eq:E33}, increases monotonically as the redshift decreases, since $z \downarrow \Rightarrow\ (1 + z)^{-2.5} \uparrow \Rightarrow\ \kappa(z) \uparrow$, and the exponent $\xi(z)$ is likewise an increasing or only weakly varying function of redshift; hence $\varphi(z)$ increases as $z$ decreases. For any positive $\varphi$, the right-hand side of Equation~\eqref{eq:C1} is a rational function of $\varphi$ whose derivative is strictly positive,
\begin{linenomath}
\begin{equation}\label{eq:C2}
\frac{d}{d\varphi}\left(\frac{\Omega_{\mathrm{vol}}\,\varphi}{\Omega_{m} + \Omega_{\mathrm{vol}}\,\varphi}\right) = \frac{\Omega_{\mathrm{vol}}\,\Omega_{m}}{(\Omega_{m} + \Omega_{\mathrm{vol}}\,\varphi)^{2}} > 0,
\end{equation}
\end{linenomath}
because $\Omega_{\mathrm{vol}}$, $\Omega_{m}$ and the squared denominator are all positive. Consequently $\Omega_{\Lambda,\mathrm{het}}$ increases strictly monotonically with $\varphi$, and since $\varphi$ grows as $z$ decreases,
\begin{linenomath}
\begin{equation}\label{eq:C3}
\frac{d\Omega_{\Lambda,\mathrm{het}}}{dz} < 0, \qquad \frac{d\Omega_{\Lambda,\mathrm{het}}}{dt} > 0.
\end{equation}
\end{linenomath}
The dark-energy density parameter therefore increases with cosmic time toward the present, even though the same $z$-dependent term $\varphi(z)$ appears in both the numerator and the denominator of Equation~\eqref{eq:C1}; the constant matter term $\Omega_{m}$ in the denominator is what makes the fraction grow with $\varphi$ and approach unity asymptotically. This asymptotic limit is not attained within the parameter range considered, because $\kappa(z)$ and $\xi(z)$ remain small as $z \to 0$ and an approach to unity would require redshifts in the unphysical negative domain. The variation of $\Omega_{\Lambda,\mathrm{het}}(z)$ is therefore confined to the interval $[\approx 0.68; 0.735]$ over $z \in [10^{3}; 10^{-4}]$, consistent with Figure~\ref{fig:OmegaLambda}.

\section[\appendixname~\thesection]{\texorpdfstring{$t_{U}$, $H_{0}$ and $\sigma_{8}$}{tU, H0 and sigma8} as Effective Averaged Quantities, and Outlook}\label{app:effective}

When the cosmic-clock constraint of Tomasetti et al. \cite{Tomasetti2026}, $t_{U} \geq 13.8 \pm 1.0\,\text{(stat.)} \pm 1.4\,\text{(syst.)}$\,Gyr, is interpreted within a homogeneous $\Lambda$CDM framework, the inferred cosmic age is related to an upper limit on the Hubble constant $H_{0}$, because the age depends directly on the assumed expansion history. That interpretation, however, rests on a globally homogeneous universe, in which the expansion is described by the Friedmann--Lemaître dynamics and a redshift-dependent Hubble parameter $H(z)$ derived from a single present-day $H_{0}$. The standard model does not explicitly include the dynamical influence of large-scale heterogeneity on the evolution of cosmic time itself. In this sense, the observationally inferred quantities $t_{U}$, $H_{0}$ and $\sigma_{8}$ may represent effective or averaged values when they are interpreted within a homogeneous cosmological model.

Once heterogeneity is included as a dynamical component, as in Section~\ref{sec:time}, the relation between redshift, expansion, dark energy and cosmic time is modified: the Hubble parameter, the Hubble constant $H_{0}(z)$ and the dark-energy density parameter $\Omega_{\Lambda}(z)$ become coupled to the time evolution of the heterogeneous Universe. The cosmic age is then no longer an invariant quantity determined solely by the homogeneous $\Lambda$CDM expansion history, but depends on the model used to describe the real, structured Universe. The self-consistent age of the heterogeneous Universe, $t_{U} = 13.521^{+0.355}_{-0.338}$\,Gyr, remains compatible with the cosmic-clock constraint and therefore does not contradict the stellar-age lower bound.

These results suggest that the cosmic-clock age constraint, the observed discrepancy in $H_{0}$ and the time evolution of $\Omega_{\Lambda}(z)$ may be different observational manifestations of the same underlying fact, namely that the Universe is not perfectly homogeneous and that its heterogeneity influences the progression of cosmic time. The functional relations and numerical evaluations derived in this work may serve as a basis for future investigations, including more precise numerical simulations, further theoretical refinements and future empirical tests through cosmological observations across a broad range of redshift.

\section[\appendixname~\thesection]{Step-by-Step Derivations}\label{app:derivations}

The methods presented here derive essential physical results from the volume dynamics of space, including the Schrödinger equation, the quantum postulates, gravity and the curvature of spacetime.

\noindent\textbf{Heterogeneous Hubble constant and dark-energy density.} The functions $H_{0,\mathrm{het}}(z)$ [Equation~\eqref{eq:H0het}] and $\Omega_{\Lambda,\mathrm{het}}(z)$ [Equation~\eqref{eq:OLhet}] are derived along the chain summarized in the derivation map (Appendix~\ref{app:map}). The most insightful derivation steps are presented next.

\noindent(1) At each point $P$ in the Universe, the relativistic energy is
\begin{linenomath}
\begin{equation}\label{eq:E1}
E(v) = E_{0}/\sqrt{1 - \vec{v}^{2}/c^{2}}.
\end{equation}
\end{linenomath}
Here $\vec{v}$ is the velocity relative to the local \textit{adequate coordinate system} (ACS), which exists at each $P$ \cite{Carmesin2026}; relative to any unspecific coordinate system, the velocity of the ACS is uniquely determined.

\noindent(2) Schwarzschild \cite{Schwarzschild1916} showed that, at a distance $r$ from a mass $M$, space is stretched in the radial direction by the factor $1/\sqrt{1 - R_{S}/r}$, with the Schwarzschild radius $R_{S} = 2GM/c^{2}$. Consequently, additional volume portions exist near $M$, and since $M$ in general moves, these portions move with it.

\noindent(3) To derive a fundamental theory, space and its fundamental volume portions are considered. The energy $\delta E$ of each volume portion $\delta V$ is the dark-energy density $u_{\Lambda}$ times $\delta V$ and is therefore positive, while $\delta V$ carries no rest mass \cite{PDG2024}. Equation~\eqref{eq:E1} with $m_{0,\mathrm{vol}} = 0$ implies that each volume portion moves at the speed of light $c$ \cite{Carmesin2026}.

\noindent(4) Space cannot be a single entity. Since the energy density of space is positive, its energy $E$ is non-zero, so that Equation~\eqref{eq:E1} may be divided by $E^{2}$. With the vanishing rest energy of a volume portion from step (3) this gives
\begin{linenomath}
\begin{equation}\label{eq:E2}
1 - \frac{\vec{v}^{\,2}}{c^{2}} = \frac{E_{0}^{2}}{E^{2}} = 0, \qquad \text{hence} \qquad v = c.
\end{equation}
\end{linenomath}
A single entity would therefore move at the speed of light parallel to one direction $\vec{e}_{v}$, which contradicts the isotropy of space observed on large scales \cite{Planck2020}. This contradiction is the \textit{space paradox}. Four of its five premises are established by observation: the isotropy of space, the positive energy density, the energy--speed relation in the adequate coordinate system, and the vanishing rest energy of volume. The fifth, that space is a single entity, is a convention of the usual description rather than a result, and it is the premise that has to be given up. Space therefore consists of parts $\delta V$, each moving at speed $c$, whose velocity directions are distributed isotropically and average to zero. Isotropy is thereby restored, and space is a stochastic average of fundamental volume portions \cite{Carmesin2026}.

\noindent(5) These volume portions are indivisible. Assume that a portion $\delta V_{j}$ with $v_{j} = c$ consisted of smaller parts $\delta V_{k}$. Each part would again have speed $c$, hence a velocity $c\,\vec{e}_{k}$ with a unit vector $\vec{e}_{k}$. In a homogeneous universe no source selects a common direction for the $\vec{e}_{k}$, so the velocity of $\delta V_{j}$ would be the average of the $\vec{v}_{k}$ and would have an absolute value below $c$, contradicting $v_{j} = c$. The assumption therefore fails, and each fundamental volume portion is indivisible \cite{Carmesin2026}. This indivisibility is what makes the volume portions quantized, and it is the property on which step (10) rests.

\noindent(6) To obtain a general, mathematically founded rule for the motion and form of each volume portion, $\delta V$ is described by a function $\varepsilon_{L}$ of space and time. It is convenient, for instance in the Schwarzschild metric, to use two distance measures: a light-travel distance $\vec{L}$, $\tau$ describing the real curved spacetime, and a gravitational-parallax distance $\vec{R}$, $t$ describing the underlying measurable flat spacetime \cite{Carmesin2026}. Near a mass $M$ the radial difference $\Delta R$ is stretched to the light-travel difference $\Delta L$, while the two transverse differences are unchanged,
\begin{linenomath}
\begin{equation}\label{eq:E3}
\Delta L = \Delta R\,\sqrt{g_{RR}}, \quad g_{RR} = \frac{1}{1 - R_{S}/R}, \quad R_{S} = \frac{2GM}{c^{2}},
\end{equation}
\end{linenomath}
so that the corresponding volumes obey $\Delta V_{L} = \Delta V_{R}\sqrt{g_{RR}}$. Each additional volume portion $\delta V := \Delta V_{L} - \Delta V_{R}$ therefore exists in addition to an already present volume $\Delta V_{R}$, and the relative additional volume is
\begin{linenomath}
\begin{equation}\label{eq:E4}
\varepsilon_{L} := \frac{\delta V}{\Delta V_{L}} = 1 - \frac{\Delta V_{R}}{\Delta V_{L}} = 1 - \frac{1}{\sqrt{g_{RR}}}.
\end{equation}
\end{linenomath}
Curvature can thus be described equivalently by the metric tensor or by the relative additional volume. The second description, unlike the first, refers to volume portions and therefore admits a dynamics of these portions. The position of $\delta V$ is a local maximum of $\varepsilon_{L}(\vec{L}, \tau)$, where all changes add to zero,
\begin{linenomath}
\begin{equation*}
\frac{\partial \varepsilon_{L}}{\partial \tau}\,d\tau + \frac{\partial \varepsilon_{L}}{\partial \vec{L}} \cdot d\vec{L} = 0.
\end{equation*}
\end{linenomath}
With $d\vec{L} = c\,\vec{e}_{v}\,d\tau$ and division by $d\tau$, the dynamics is the volume-dynamics Equation~\eqref{eq:vd}. Squaring it gives the Lorentz-invariant form
\begin{linenomath}
\begin{equation}\label{eq:E5}
\left(\frac{\partial \varepsilon_{L}}{\partial \tau}\right)^{2} - c^{2}\left(\frac{\partial \varepsilon_{L}}{\partial \vec{L}}\right)^{2} = 0.
\end{equation}
\end{linenomath}
Only the existence of a local maximum and elementary calculus enter here, and no physical assumption is used. This is the fundamental equation of the dynamics, called the \textit{volume dynamics} (VD) \cite{Carmesin2026}. The contrast with the established descriptions is the decisive point: general relativity postulates the field equation and quantum physics postulates its axioms, whereas both follow below from Equation~\eqref{eq:vd}.

\noindent(7) The VD in Equation~\eqref{eq:vd} implies the Schrödinger equation \cite{Schrodinger1926}. Multiplying Equation~\eqref{eq:vd} by $i\hbar$, applying the time derivative and writing $\dot{\varepsilon}_{L}$ for the time derivative of $\varepsilon_{L}$ gives
\begin{linenomath}
\begin{equation*}
i\hbar\frac{\partial}{\partial \tau}\dot{\varepsilon}_{L} + i\hbar\frac{\partial}{\partial \vec{L}}\dot{\varepsilon}_{L} \cdot c \cdot \vec{e}_{v} = 0.
\end{equation*}
\end{linenomath}
With the momentum operator $\hat{\vec{p}} = -i\hbar\,\partial/\partial \vec{L}$, so that $\hat{\vec{p}} \cdot \vec{e}_{v} = \hat{p}$, and the energy $\hat{E} = \hat{p}\,c$ for an object of speed $c$, identifying the normalized rate $t_{n}\dot{\varepsilon}_{L}$ with the wave function $\Psi$ yields the Schrödinger equation
\begin{linenomath}
\begin{equation}\label{eq:E6}
i\hbar\frac{\partial}{\partial \tau}\Psi = \hat{E} \cdot \Psi.
\end{equation}
\end{linenomath}
Here $t_{n}$ is a normalization of dimension time, which makes $\Psi$ dimensionless. Since $\dot{\varepsilon}_{L}$ is the derivative of a real quantity, the wave function obtained in this way is real. Equation~\eqref{eq:E6} holds for objects of arbitrary speed and is therefore a generalized Schrödinger equation. For a slow massive object, $p^{2}c^{2} \ll m_{0}^{2}c^{4}$, the energy is $E \doteq E_{0} + p^{2}/(2m_{0})$; inserting this, substituting the operator $\hat{p}$ for $p$ and factorizing $\Psi_{E_{0}} = \Psi\exp(E_{0}\tau/i\hbar)$, the rest-energy terms cancel and the ordinary Schrödinger equation follows,
\begin{linenomath}
\begin{equation}\label{eq:E7}
i\hbar\frac{\partial}{\partial \tau}\Psi \doteq \frac{\hat{p}^{2}}{2m_{0}}\Psi + E_{pot}\,\Psi = \hat{H}\Psi.
\end{equation}
\end{linenomath}
The fundamental volume portions thus obey the Schrödinger equation \cite{Carmesin2026}. By the Higgs mechanism \cite{Higgs1964}, mass forms from volume portions through a phase transition, so the same dynamics holds on both sides of the transition and the Schrödinger equation holds for matter as well; the remaining quantum postulates \cite{Hilbert1928,Kumar2018} have likewise been derived from the VD \cite{Carmesin2026}.

\noindent(8) To obtain gravity from the VD, Equation~\eqref{eq:vd} is multiplied by $c$:
\begin{linenomath}
\begin{equation}\label{eq:E8}
c\frac{\partial \varepsilon_{L}}{\partial \tau} + \left(-\frac{\partial}{\partial \vec{L}}\big[\,\underbrace{-c^{2}\varepsilon_{L}}_{\Phi_{L}}\,\big]\right) \cdot \vec{e}_{v} = 0.
\end{equation}
\end{linenomath}
Applied to the indivisible volume portions in a neighborhood $S$ of an event, the bracketed term becomes a generalized potential,
\begin{linenomath}
\begin{equation}\label{eq:E9}
\Phi_{gen}(\tau, \vec{L}) := -c^{2}\sum_{j \in S}\varepsilon_{L,rr,j},
\end{equation}
\end{linenomath}
and its negative gradient is the corresponding field. The summation is admissible because a volume portion carries a change of quadrupolar structure, hence a tensor of rank two and an integer spin, so that the portions are bosons and the relative additional volumes caused by different masses add at the same point. The bracketed term is in fact an exact gravitational potential \cite{Carmesin2026}. Applying the gradient gives the gravitational field $\vec{G}^{*}$, with absolute value $G^{*}$
\begin{linenomath}
\begin{equation}\label{eq:E10}
c\frac{\partial \varepsilon_{L}}{\partial \tau} + G^{*} = 0, \qquad \vec{G}^{*} = \frac{GM}{R^{2}}\,\vec{e}_{v} \quad \text{near a mass.}
\end{equation}
\end{linenomath}
The inverse-square dependence in Equation~\eqref{eq:E10} is not assumed but follows. Integrating Equation~\eqref{eq:E10} over a short time $\underline{\delta}\tau$ gives $\vec{e}_{v}\,c\sum_{j}\varepsilon_{L,rr,j} = -\underline{\delta}\tau\,\vec{G}^{*}$. With $\varepsilon_{L,rr,j} = \delta V_{j}/\Delta V_{L}$ and, since each portion propagates at speed $c$ and carries the positive energy density $u_{vol}$, with $\delta V_{j} = c\,\delta p_{j}/u_{vol}$, and with a shell $\Delta V_{L} = 4\pi R^{2}\,\Delta L$ of constant thickness $\Delta L$ centered on the mass,
\begin{linenomath}
\begin{equation}\label{eq:E11}
|\vec{G}^{*}| = \frac{c^{2}}{u_{vol}\,4\pi\,\Delta L} \cdot \frac{\sum_{j}\delta p_{j}}{\underline{\delta}\tau} \cdot \frac{1}{R^{2}}.
\end{equation}
\end{linenomath}
The middle factor is the momentum current passing through the shell. No shell creates or absorbs momentum, so this current is the same for all shells, and $|\vec{G}^{*}| \propto 1/R^{2}$ follows. With the equivalence principle \cite{Galilei1638,Einstein1911} the source strength is proportional to $M$, which gives the field in Equation~\eqref{eq:E10}.

Curvature follows from the same potential. With the position factor $\varepsilon_{E} := dR/dL = 1 - \varepsilon_{L,rr}$ and $\Phi_{L} = -c^{2}\varepsilon_{L,rr}$, applying $\partial/\partial L$ and identifying the field gives
\begin{linenomath}
\begin{equation}\label{eq:E12}
\varepsilon_{E}(R)\,\frac{\partial \varepsilon_{E}}{\partial R} = \frac{|\vec{G}^{*}|}{c^{2}} = \frac{GM}{R^{2}c^{2}}.
\end{equation}
\end{linenomath}
Integration, with the boundary condition $\varepsilon_{E} \to 1$ for $R \to \infty$, which expresses the absence of curvature at large distance, yields
\begin{linenomath}
\begin{equation}\label{eq:E13}
\varepsilon_{E} = \sqrt{1 - \frac{2GM}{c^{2}R}} = \frac{1}{\sqrt{g_{RR}}}.
\end{equation}
\end{linenomath}
The Schwarzschild form used in Equation~\eqref{eq:E3} is thereby recovered from the volume dynamics instead of being assumed, and the comparison identifies the constant of proportionality in Equation~\eqref{eq:E10} with the gravitational constant $G$. The VD therefore implies gravity as a function of $R$, the curvature of spacetime and the energy density $u_{f}$ of the gravitational field, without approximation \cite{Carmesin2026}. The result is exact because the adequate coordinate system removes the velocity-dependent terms of the post-Newtonian expansion and because the two distance measures of step (6) are kept apart throughout.

\noindent(9) Near a mass $M$ there occurs the additional volume $\delta V = \Delta V_{L}\,(1 - \Delta V_{R}/\Delta V_{L})$ (cf. step 6). For a shell of radius $R$ and thickness $\Delta R$, one has $\Delta V_{R} = 4\pi R^{2}\Delta R$ and $\Delta V_{L} = 4\pi R^{2}\Delta L$ with $\Delta L = \Delta R/\sqrt{1 - R_{S}/R}$, so that $\delta V = 4\pi R^{2}\Delta R\,(\varepsilon_{E}^{-1} - 1)$. In the far-distance approximation $R_{S}/R \ll 1$, at first order $\varepsilon_{E}^{-1} \doteq 1 + R_{S}/(2R)$, and therefore
\begin{linenomath}
\begin{equation}\label{eq:E14}
\frac{\partial}{\partial R}\,\delta V = \frac{\partial}{\partial R}\left(4\pi R^{2}\Delta R \cdot \frac{R_{S}}{2R}\right) = 2\pi\,\Delta R\,R_{S}.
\end{equation}
\end{linenomath}
The shell's volume portions move outwards at $v = c$, so during a time $\underline{\delta}\tau$ the additional volume advances by $\underline{\delta}R = c\,\underline{\delta}\tau$, and the volume formed in a shell of that thickness is $\underline{\delta}V = \underline{\delta}R\,\partial(\delta V)/\partial R = 2\pi R_{S}\,\Delta R\,\underline{\delta}R$. The change per time and per volume $\Delta V_{L}$ is
\begin{linenomath}
\begin{equation}\label{eq:E15}
\underline{\dot{\varepsilon}}_{L} := \frac{\underline{\delta}V}{\underline{\delta}\tau \cdot \Delta V_{L}} = \frac{2\pi R_{S}\,\Delta R\,\underline{\delta}R}{(\underline{\delta}R/c) \cdot 4\pi R^{2}\Delta R} = \frac{R_{S}\,c}{2R^{2}},
\end{equation}
\end{linenomath}
called the \textit{locally formed volume} (LFV). Inserting $R_{S} = 2GM/c^{2}$ gives $c\,\underline{\dot{\varepsilon}}_{L} = GM/R^{2} = G^{*}$. In general, a gravitational field $\vec{G}^{*}$ causes the radial LFV rate $\underline{\dot{\varepsilon}}_{L,rr}$, which gives the \textit{law of LFV} \cite{Carmesin2026}:
\begin{linenomath}
\begin{equation}\label{eq:E16}
\underline{\dot{\varepsilon}}_{L,rr} = \frac{G^{*}}{c}.
\end{equation}
\end{linenomath}
Squaring Equation~\eqref{eq:E16} shows that it is the square of the field that determines the rate. This is the step on which the treatment of heterogeneity in the main text rests, because a field distribution with a vanishing mean can have a non-vanishing mean square.

\noindent(10) For the early Universe ($z \approx 1000$), the law of LFV in Equation~\eqref{eq:E16} yields the dark-energy density $u_{vol} = \lim_{z \to \infty} u_{\Lambda} = \rho_{vol}c^{2}$. The derivation uses an empty probe volume $dV_{0}$ and collects the contributions that reach it. Each mass element $dM_{j}$ in a shell of radius $R$ around $dV_{0}$ causes there, by Equation~\eqref{eq:E16}, the rate
\begin{linenomath}
\begin{equation}\label{eq:E17}
d\underline{\dot{\varepsilon}}_{L,rr} = \frac{|d\vec{G}^{*}_{j}|}{c} = \frac{G\,dM_{j}}{R^{2}c}.
\end{equation}
\end{linenomath}
Summing over the shell with $dM = \rho_{vol}\,4\pi R^{2}\,dR$, the factor $R^{2}$ cancels, and with the light-travel time $dt_{LT} = dR/c$ of a volume portion crossing the shell,
\begin{linenomath}
\begin{equation}\label{eq:E18}
d\underline{\dot{\varepsilon}}_{L,rr} = 4\pi G\,\rho_{vol}\,dt_{LT}.
\end{equation}
\end{linenomath}
The shells that can contribute are exactly those whose volume portions reach $dV_{0}$, that is, those with light-travel times between zero and the Hubble time $t_{H_{0}}$. The spatial integration is therefore carried out over $t_{LT}$ and gives
\begin{linenomath}
\begin{equation}\label{eq:E19}
\underline{\dot{\varepsilon}}_{L,rr} = 4\pi G\,\rho_{vol}\,t_{H_{0}} = H_{0},
\end{equation}
\end{linenomath}
where the rate at $dV_{0}$ is identified with the Hubble constant because $H_{0} = \dot{V}/(3V) = \underline{\dot{\varepsilon}}_{L,iso}/3 = \underline{\dot{\varepsilon}}_{L,rr}$. Equation~\eqref{eq:E19} also provides a test of completeness: the volume formed at $dV_{0}$ during one Hubble time is
\begin{linenomath}
\begin{equation}\label{eq:E20}
\underline{\delta}V = \underline{\dot{\varepsilon}}_{L,rr}\,t_{H_{0}}\,dV_{0} = H_{0}\,t_{H_{0}}\,dV_{0} = dV_{0},
\end{equation}
\end{linenomath}
so the local process forms exactly the probe volume, and by translation invariance exactly the volume of the Universe. With $t_{H_{0}} = 1/H_{0}$, Equation~\eqref{eq:E19} gives the dark-energy density
\begin{linenomath}
\begin{equation}\label{eq:E21}
\rho_{vol} = \frac{H_{0}^{2}}{4\pi G}, \qquad u_{vol} = \frac{H_{0}^{2}c^{2}}{4\pi G} = (5.034 \pm 0.14) \times 10^{-10}\ \mathrm{J}/\mathrm{m}^{3},
\end{equation}
\end{linenomath}
in agreement, within the stated uncertainties, with the value $u_{vol,\mathrm{obs}} = (5.133 \pm 0.24) \times 10^{-10}\ \mathrm{J}/\mathrm{m}^{3}$ observed at the cosmic microwave background ($z_{CMB} = 1090.30 \pm 0.41$) \cite{Planck2020}. Dividing by the critical density $\rho_{cr.,0} = 3H_{0}^{2}/(8\pi G)$ removes the epoch of observation and leaves a pure number,
\begin{linenomath}
\begin{equation}\label{eq:E22}
\Omega_{vol} = \frac{\rho_{vol}}{\rho_{cr.,0}} = \frac{H_{0}^{2}/(4\pi G)}{3H_{0}^{2}/(8\pi G)} = \frac{2}{3},
\end{equation}
\end{linenomath}
matching the observed $\Omega_{\Lambda}(z_{CMB}) = 0.679 \pm 0.013$ \cite{Planck2020} within one standard deviation. This is the value used as $\Omega_{vol}$ throughout the main text.

A homogeneous matter or radiation density contributes nothing to the rate at $dV_{0}$. Its gravitational fields are classical, and both their mean and the mean of their squares vanish when averaged over the shells, so no LFV is caused there. For the indivisible volume portions the mean field vanishes but the mean square does not, because the portions are quantized; the same holds for the heterogeneous part of the matter density. Only these two therefore drive LFV at $dV_{0}$, and the LFV they cause does not average to zero \cite{Carmesin2026}. In a heterogeneous universe, the heterogeneity produces a vanishing mean field $\langle \vec{G}^{*} \rangle$ but a non-zero mean square $\langle (\vec{G}^{*})^{2} \rangle$, which causes LFV at $dV_{0}$. As $z$ decreases the heterogeneity grows, the LFV increases, and the resulting Hubble constant $H_{0}(z)$ increases; this function reproduces observations at many redshifts and explains the Hubble tension as a consequence of the unifying derivation summarized in the derivation map (Appendix~\ref{app:map}) \cite{Carmesin2026}.

(11) In the heterogeneous Universe, for each redshift $z$, the derived Hubble constant is Equation~\eqref{eq:H0het}, which follows from \cite{Carmesin2026} with the dark-energy density $\rho_{vol} = H_{0}^{2}/(4\pi G)$ and the auxiliary functions $\kappa(z)$ and $\xi(z)$. We identify the two summands in its root by $\rho_{m,\mathrm{het}}$ and $\rho_{\Lambda,\mathrm{het}}$, and their sum by the critical density of the heterogeneous Universe, $\rho_{cr.,\mathrm{het}}$, since the curvature parameter vanishes in Equation~\eqref{eq:H0het}:
\begin{linenomath}
\begin{equation}\label{eq:E23}
\rho_{cr.,\mathrm{het}} = \rho_{m,\mathrm{het}} + \rho_{\Lambda,\mathrm{het}}.
\end{equation}
\end{linenomath}
Applying the usual density parameters,
\begin{linenomath}
\begin{equation}\label{eq:E24}
\Omega_{j} := \frac{\rho_{j}}{\rho_{cr.,0}}; \qquad \Omega_{j,\mathrm{het}} := \frac{\rho_{j,\mathrm{het}}}{\rho_{cr.,\mathrm{het}}},
\end{equation}
\end{linenomath}
and expanding Equation~\eqref{eq:H0het} by $\sqrt{\rho_{cr.,\mathrm{het}}}$ gives
\begin{linenomath}
\begin{equation}\label{eq:E25}
H_{0,\mathrm{het}} = H_{0,\Lambda\mathrm{CDM}} \cdot \sqrt{\underbrace{\frac{\rho_{cr.,\mathrm{het}}}{\rho_{cr.,0}}}_{\Omega_{cr.,\mathrm{het}}}} \cdot \sqrt{\Omega_{m,\mathrm{het}} + \Omega_{\Lambda,\mathrm{het}}}.
\end{equation}
\end{linenomath}
With $\Omega_{cr.,\mathrm{het}} = \rho_{cr.,\mathrm{het}}/\rho_{cr.,0}$ from Equation~\eqref{eq:E24},
\begin{linenomath}
\begin{equation}\label{eq:E26}
\Omega_{cr.,\mathrm{het}} = \frac{\rho_{cr.,\mathrm{het}}}{\rho_{cr.,0}},
\end{equation}
\end{linenomath}
and, using Equations~\eqref{eq:E23} and \eqref{eq:E24},
\begin{linenomath}
\begin{equation}\label{eq:E27}
\Omega_{m,\mathrm{het}} = \frac{\rho_{m}}{\rho_{cr.,\mathrm{het}}} = \frac{\Omega_{m}}{\Omega_{m} + \Omega_{vol} \cdot (1 + \kappa)^{\xi}},
\end{equation}
\end{linenomath}
which, with $\rho_{vol} = H_{0}^{2}/(4\pi G)$, yields the dark-energy density parameter $\Omega_{\Lambda,\mathrm{het}}$, Equation~\eqref{eq:OLhet}. Both parameters are normalized to the same critical density, so that they add up to unity: the two fractions in Equations~\eqref{eq:E27} and \eqref{eq:OLhet} share the common denominator $\Omega_{m} + \Omega_{\mathrm{vol}}(1 + \kappa)^{\xi}$, while their numerators are $\Omega_{m}$ and $\Omega_{\mathrm{vol}}(1 + \kappa)^{\xi}$, whence $\Omega_{m,\mathrm{het}} + \Omega_{\Lambda,\mathrm{het}} = 1$ exactly, as required for the proven spatially flat Universe \cite{Carmesin2023a,Carmesin2023b}. At the present day this gives $\Omega_{m,\mathrm{het}}(x_{0}) = 0.2649$ together with $\Omega_{\Lambda,\mathrm{het}}(x_{0}) = 0.7351$. The radiation term is negligible in this balance: relevant structure formation, and hence the growth of heterogeneity, sets in long after the radiation era, so that including $\Omega_{r}$ in the normalization would change the results only in the fifth decimal place, well below the quoted uncertainties.

The present-day value $\Omega_{m,\mathrm{het}}(x_{0}) = 0.2649$ connects the observed ranges systematically. At high redshift the heterogeneity vanishes, $\kappa \to 0$, the denominator $\Omega_{m} + \Omega_{\mathrm{vol}}(1 + \kappa)^{\xi}$ approaches unity, and $\Omega_{m,\mathrm{het}}$ approaches the Planck value $\Omega_{m} = 0.321$ exactly. Toward the present, $\Omega_{m,\mathrm{het}}$ decreases strictly monotonically to 0.2649, because the heterogeneity grows; accordingly, a larger $H_{0}$ corresponds to a more heterogeneous Universe and hence to a smaller $\Omega_{m,\mathrm{het}}$. This range connects to the low-redshift determinations underlying Table~\ref{tab:OmegaLambda}, which yield matter parameters of 0.25 to 0.29 for spatial flatness \cite{Perlmutter1999,Knop2003,Ghirardini2024}. The Planck determination itself rests on CMB data, which do not yet include the heterogeneous contribution and therefore render $\Omega_{m,\mathrm{het}}$ only approximately. Such a monotonic connection of the early- and late-Universe ranges is not provided by a free fit.

\noindent\textbf{Cosmic age from the integrated expansion history.} The age $t_{U}(z)$ of the Universe is derived as a function of the redshift $z$ for the homogeneous and the heterogeneous case. As a method, the relation between the increments $da$ of the scale radius and $dt$ of the time is used, see, e.g., \cite{Hobson2006}:
\begin{adjustwidth}{-\extralength}{0cm}
\begin{linenomath}
\begin{equation}\label{eq:E28}
\int_{a_{1}}^{a_{2}} \frac{x \cdot da/a_{0}}{H_{0} \cdot \sqrt{\Omega_{r} + \Omega_{m} \cdot x + \Omega_{\Lambda} \cdot x^{4}}} = \int_{t_{1}}^{t_{2}} dt' \quad \text{or} \quad \frac{x \cdot da/a_{0}}{H_{0} \cdot \sqrt{\Omega_{r} + \Omega_{m} \cdot x + \Omega_{\Lambda} \cdot x^{4}}} = dt.
\end{equation}
\end{linenomath}
\end{adjustwidth}
Hereby, $\Omega_{r} = 9.265 \times 10^{-5}$ is the density parameter of radiation, as implied by the Planck parameters \cite{Planck2020}, $a_{0}$ is the present-day value of the scale radius, $x = a/a_{0}$ is a scaled scale radius, and $x$ is the following function of the redshift:
\begin{linenomath}
\begin{equation}\label{eq:E29}
x = \frac{1}{1 + z} \quad \text{or} \quad z = \frac{1}{x} - 1.
\end{equation}
\end{linenomath}
The overall procedure follows three consecutive steps, each of which requires only the result of the previous one, so that the three problems are solved sequentially rather than simultaneously; no feedback occurs at leading order in the added heterogeneity, in which the effects are evaluated here, while higher orders could be determined systematically. In the first step, carried out in \cite{Carmesin2026}, the rate of the expansion is obtained: the standard deviation of the gravitational field strengths caused by heterogeneity yields the additional rate of locally formed volume and thus $H_{0,\mathrm{het}}(z)$, Equation~\eqref{eq:E32}; the normalization of the density parameters does not yet enter, because no integration is performed. In the second step, with the rate known, the normalization is determined: the critical density of the heterogeneous Universe, Equation~\eqref{eq:E23}, provides $\Omega_{m,\mathrm{het}}$ and $\Omega_{\Lambda,\mathrm{het}}$, Equations~\eqref{eq:E27} and \eqref{eq:OLhet}. In the third step, with rate and normalization at hand, the three quantities $H_{0}$, $\Omega_{m}$ and $\Omega_{\Lambda}$ are replaced by their heterogeneous counterparts in the time increment $dt$, and the integration yields the cosmic time. The sequence must begin with the rate, because the volume dynamics directly provides it. This sequence also bears on the timescape scenario \cite{Wiltshire2007}, which demonstrates that observations can be described with or without a cosmological constant and thus leaves the value of $\Omega_{\Lambda}$ open; the derivation of $\Omega_{\mathrm{vol}} = 2/3$ from the formation process of volume decides this question without requiring a specific model of dark energy.

\begin{figure}[!t]
\includegraphics[width=8.5cm]{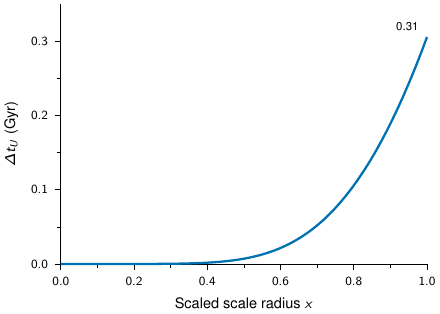}
\caption{The global time difference $\Delta t_{U} = t_{U,\mathrm{hom}} - t_{U,\mathrm{het}}$ as a function of the scaled scale radius $x$.\label{fig:DeltatUx}}
\end{figure}

As the Hubble constant is a function of the redshift, see Eq.~(244) in \cite{Carmesin2026} and Equation~\eqref{eq:E32} below, $H_{0}$ is a function of the redshift $z$ or the scaled scale radius $x$. Similarly, the density parameter of dark energy is a function of $x$, see Equation~\eqref{eq:OLhet}. Additionally, $da/a_{0} = dx$. Therefore, $dt$ in Equation~\eqref{eq:E28} becomes $dt_{\mathrm{het}}$ in the heterogeneous Universe, when $H_{0}$ becomes $H_{0,\mathrm{het}}$, $\Omega_{\Lambda}$ becomes $\Omega_{\Lambda,\mathrm{het}}$ and $\Omega_{m}$ becomes $\Omega_{m,\mathrm{het}}$, Equation~\eqref{eq:E27}; in the heterogeneous Universe all quantities are heterogeneous, and consequently so is the increment of time:
\begin{linenomath}
\begin{equation}\label{eq:E30}
\frac{x \cdot dx}{H_{0,\mathrm{het}}(x) \cdot \sqrt{\Omega_{r} + \Omega_{m,\mathrm{het}}(x) \cdot x + \Omega_{\Lambda,\mathrm{het}}(x) \cdot x^{4}}} = dt_{\mathrm{het}}.
\end{equation}
\end{linenomath}
In order to derive the age of the heterogeneous Universe $t_{U,\mathrm{het}}(x)$ as a function of the scaled scale radius $x$, the above equation is integrated. The integral starts at $x = 0$ and $t_{U,\mathrm{het}}(x = 0) = 0$, corresponding to the Big Bang. For a chosen arbitrary upper bound $x_{2}$, the integral has a corresponding time $t_{U,\mathrm{het}}(x_{2})$:
\begin{adjustwidth}{-\extralength}{0cm}
\begin{linenomath}
\begin{equation}\label{eq:E31}
\int_{0}^{x_{2}} \frac{x \cdot dx}{H_{0,\mathrm{het}}(x) \cdot \sqrt{\Omega_{r} + \Omega_{m,\mathrm{het}}(x) \cdot x + \Omega_{\Lambda,\mathrm{het}}(x) \cdot x^{4}}} = \int_{0}^{t_{U,\mathrm{het}}(x_{2})} dt' = t_{U,\mathrm{het}}(x_{2}).
\end{equation}
\end{linenomath}
\end{adjustwidth}
Hereby, for the case of the homogeneous Universe, the density parameter $\Omega_{\Lambda,\mathrm{het}}$ takes the value $\Omega_{\mathrm{vol}} = \Omega_{\Lambda,\mathrm{hom}}$, and $\Omega_{m,\mathrm{het}}$ takes the value $\Omega_{m}$. The above integral is solved numerically. Thereby, the Hubble constant is expressed as a function of the redshift $z$ as follows, see Eq.~(244) in \cite{Carmesin2026}:
\begin{adjustwidth}{-\extralength}{0cm}
\begin{linenomath}
\begin{equation}\label{eq:E32}
H_{0,\mathrm{het}}(x) \mathrel{\widehat{=}} H_{0,\mathrm{het}}(z) = H_{0,\Lambda\mathrm{CDM}}\sqrt{\Omega_{m} + \Omega_{\mathrm{vol}} \cdot (1 + \kappa(z))^{\xi(z)}}, \quad \text{with} \quad z = \frac{1}{x} - 1.
\end{equation}
\end{linenomath}
\end{adjustwidth}
Hereby, $\kappa(z)$, see Eq.~(223) in \cite{Carmesin2026}, is used in a meaningful approximation at first order, $q_{t}^{(n)} = q_{1}^{(1)}$, so that:
\begin{linenomath}
\begin{equation}\label{eq:E33}
\kappa(z) = \frac{\Omega_{m} \cdot \sigma_{8}}{2\,\Omega_{\mathrm{vol}} \cdot (1 + z)^{2.5}} = \frac{\Omega_{m} \cdot \sigma_{8}}{2\,\Omega_{\mathrm{vol}}} \cdot x^{2.5}.
\end{equation}
\end{linenomath}
Additionally, the exponent is obtained from Eqs.~(237--240) in \cite{Carmesin2026}:
\begin{linenomath}
\begin{equation}\label{eq:E34}
y \coloneqq 1 + \kappa, \quad w_{+} = \frac{\Omega_{\mathrm{vol}} \cdot y^{2}}{2}\left(1 + \sqrt{1 + \frac{4\Omega_{m}}{\Omega_{\mathrm{vol}}^{2} \cdot y^{2}}}\right), \quad \xi = \frac{\ln(w_{+})}{\ln(y)}.
\end{equation}
\end{linenomath}
Hereby, $w_{+}$ is the physical solution for the combination $(1 + \kappa)^{\xi}$. Moreover, in the above integral, Equation~\eqref{eq:E31}, Equation~\eqref{eq:OLhet} provides the density parameter $\Omega_{\Lambda,\mathrm{het}}$ of dark energy as a function of $x = 1/(z + 1)$, whereby $\kappa$ and $\xi$ are provided by Equations~\eqref{eq:E33} and \eqref{eq:E34}. As a method of numerical integration, for a chosen increment $\Delta x$, and for a chosen value $x_{2}$, the integral in Equation~\eqref{eq:E31} is expressed by the following sum:
\begin{adjustwidth}{-\extralength}{0cm}
\begin{linenomath}
\begin{equation}\label{eq:E35}
t_{U,\mathrm{het}}(x_{2}) = \sum_{i=1}^{i_{2}} \frac{x \cdot \Delta x}{H_{0,\mathrm{het}}(x) \cdot \sqrt{\Omega_{r} + \Omega_{m,\mathrm{het}}(x) \cdot x + \Omega_{\Lambda,\mathrm{het}}(x) \cdot x^{4}}}, \quad x = i \cdot \Delta x, \quad x_{2} = i_{2} \cdot \Delta x.
\end{equation}
\end{linenomath}
\end{adjustwidth}
The above sum is evaluated numerically with the following cosmological parameters. In the numerical investigation, the CMB values in Equation~\eqref{eq:E36} are used. Therefore, $\Omega_{\mathrm{vol}} = 2/3$ is replaced by $\Omega_{\Lambda} = 0.679$. Thereby, the values in Tab.~2, column 2, of \cite{Planck2020} are utilized, as these values are especially pure and unmixed:
\begin{adjustwidth}{-\extralength}{0cm}
\begin{linenomath}
\begin{equation}\label{eq:E36}
\begin{alignedat}[b]{3}
\Omega_{\mathrm{vol}} &= \frac{2}{3}, &\qquad \Omega_{\Lambda} &= 0.679 \pm 0.013, &\qquad \sigma_{8} &= 0.8118 \pm 0.0089,\\
H_{0,\Lambda\mathrm{CDM}} &= (66.88 \pm 0.92)\,\frac{\mathrm{km}}{\mathrm{s} \cdot \mathrm{Mpc}}, &\qquad \Omega_{m} &= 0.321 \pm 0.013, &\qquad \Omega_{r} &= 9.265 \times 10^{-5}.
\end{alignedat}
\end{equation}
\end{linenomath}
\end{adjustwidth}
Evaluating this sum for the present day, $x_{2} = 1 =: x_{0}$, gives the heterogeneous age quoted in Equation~\eqref{eq:tUhet}. Similarly, the age $t_{U,\mathrm{hom}}(x_{2})$ of the homogeneous Universe is obtained with the sum
\begin{adjustwidth}{-\extralength}{0cm}
\begin{linenomath}
\begin{equation}\label{eq:E37}
t_{U,\mathrm{hom}}(x_{2}) = \sum_{i=1}^{i_{2}} \frac{x \cdot \Delta x}{H_{0,\Lambda\mathrm{CDM}} \cdot \sqrt{\Omega_{r} + \Omega_{m} \cdot x + \Omega_{\mathrm{vol}} \cdot x^{4}}}, \quad x = i \cdot \Delta x, \quad x_{2} = i_{2} \cdot \Delta x.
\end{equation}
\end{linenomath}
\end{adjustwidth}
which for the present day gives the homogeneous age, Equation~\eqref{eq:tUhom}. The resulting global time difference
\begin{linenomath}
\begin{equation}\label{eq:E38}
\Delta t_{U} = t_{U,\mathrm{hom}} - t_{U,\mathrm{het}}
\end{equation}
\end{linenomath}
is shown as a function of the redshift in Figure~\ref{fig:DeltatUz} and as a function of the scaled scale radius in Figure~\ref{fig:DeltatUx}; at the present day it amounts to $\Delta t_{U}(x_{0}) = 0.306 \pm 0.013$\,Gyr.

\noindent\textbf{Comparison with the calculation using the derived cosmological parameters.} The numerical evaluation above adopts the measured Planck values, Equation~\eqref{eq:E36}. As a complementary calculation, the same integration, including the normalization of the density parameters, is carried out with the cosmological parameters derived in the present framework: $\Omega_{\mathrm{vol}} = 2/3$ exactly, $\Omega_{k} = 0$, the radiation parameter $\Omega_{r} = \Omega_{m}/(1 + z_{\mathrm{eq}})$ with the redshift of matter-radiation equality $z_{\mathrm{eq}} = 3411 \pm 48$ from table 2, column 2, of \cite{Planck2020}, consistent with the parameter set of Equation~\eqref{eq:E36}; column 7 gives $z_{\mathrm{eq}} = 3387 \pm 21$, and the difference is negligible for $\Omega_{r}$. With $\Omega_{m} \approx 1/3$ in this step, and subsequently $\Omega_{m} = 1 - \Omega_{\Lambda} - \Omega_{r} - \Omega_{k}$. The Hubble constant $H_{0,\Lambda\mathrm{CDM}}$ and $\sigma_{8}$ remain measured inputs. The results of the two calculations are compared in Table~\ref{tab:comparison}.

\begin{table}[H]
\caption{Comparison of the two calculations: the numerical evaluation with the measured Planck values, Equation~\eqref{eq:E36}, and the same integration, including the normalization of the density parameters, carried out with the cosmological parameters derived in the present framework.\label{tab:comparison}}
\begin{tabularx}{\textwidth}{l>{\hsize=1.15\hsize}C>{\hsize=0.85\hsize}C}
\toprule
 & \textbf{Planck values, Equation~\eqref{eq:E36}} & \textbf{derived parameters}\\
\midrule
$\Omega_{\Lambda}$ resp. $\Omega_{\mathrm{vol}}$ & $0.679$ & $2/3$\\
$\Omega_{r}$ & $9.265 \times 10^{-5}$ & $9.769 \times 10^{-5}$\\
$\Omega_{m}$ & $0.321$ & $0.33324$\\
$\Omega_{\Lambda,\mathrm{het}}(x_{0})$ & $0.7351$ & $0.7265$\\
increase of $\Omega_{\Lambda,\mathrm{het}}$ & $8.3\,\%$ & $9.0\,\%$\\
$t_{U,\mathrm{het}}$ in Gyr & $13.521^{+0.355}_{-0.338}$ & $13.371^{+0.183}_{-0.178}$\\
$t_{U,\mathrm{hom}}$ in Gyr & $13.827^{+0.357}_{-0.340}$ & $13.681^{+0.191}_{-0.186}$\\
$\Delta t_{U}$ in Gyr & $0.306 \pm 0.013$ & $0.311^{+0.008}_{-0.007}$\\
\bottomrule
\end{tabularx}
\end{table}

\noindent The global time dilation is practically unchanged, $\Delta t_{U} = 0.311$ instead of 0.306~Gyr, a difference below two percent, while both absolute ages become slightly shorter. The heterogeneous evolution therefore does not depend on the adopted Planck values. Moreover, since $\Omega_{\Lambda}$ is derived rather than measured in the second calculation, its measurement uncertainty no longer contributes to the worst-case propagation, which markedly reduces the resulting uncertainties. Both spreadsheet workbooks are provided with the data set referenced under Code Availability.

\begin{adjustwidth}{-\extralength}{0cm}

\reftitle{References}

\PublishersNote{}
\end{adjustwidth}
\end{document}